\documentclass[12pt]{iopart}
\usepackage[colorlinks=true,citecolor=blue]{hyperref}
\usepackage{soul}
\usepackage{color}
\usepackage{graphicx}
\usepackage{subcaption}
\usepackage[]{graphicx}
\usepackage{booktabs}
\usepackage{multirow}
\usepackage{todonotes}

\newcommand{\argmax}{\mathop{\mathrm{arg\,max}}}

\begin{document}

\title[]{Classification of tokamak plasma confinement states with convolutional recurrent neural networks }

\author{F. Matos$^1$, V. Menkovski$^2$, F. Felici$^3$, A. Pau$^3$, F. Jenko$^1$, the TCV Team$^3$\footnote{See author list of S. Coda et al 2019 Nucl.  Fusion 59 112023} and the EUROfusion MST1 Team\footnote{See author list of B. Labit et al., 2019 Nucl. Fusion 59 086020}}

\address{$^1$ Max Planck Institute for Plasma Physics, Boltzmannstra{\ss}e 2, 85748 Garching, Germany}
\address{$^2$ Eindhoven University of Technology, 5612 AZ Eindhoven, Netherlands}
\address{$^3$ Ecole Polytechnique Fédérale de Lausanne (EPFL), Swiss Plasma Center (SPC), CH-1015 Lausanne, Switzerland}

\ead{francisco.matos@ipp.mpg.de}

\vspace{10pt}

\begin{abstract}
During a tokamak discharge, the plasma can vary between different confinement regimes: Low (L), High (H) and, in some cases, a temporary (intermediate state), called Dithering (D). In addition, while the plasma is in H mode, Edge Localized Modes (ELMs) can occur. The automatic detection of changes between these states, and of ELMs, is important for tokamak operation. Motivated by this, and by recent developments in Deep Learning (DL), we developed and compared two methods for automatic detection of the occurrence of L-D-H transitions and ELMs, applied on data from the TCV tokamak. These methods consist in a Convolutional Neural Network (CNN) and a Convolutional Long Short Term Memory Neural Network (Conv-LSTM). We measured our results with regards to ELMs using ROC curves and Youden's score index, and regarding state detection using Cohen's Kappa Index.

\end{abstract}

%
\vspace{2pc}

\normalsize

\section{Introduction}

In a fusion experiment, plasma can typically be described as being in one of two different confinement regimes or modes: Low (L) and High (H). Furthermore, the plasma can also sometimes be described as being in a third, additional, mode, called the Intermediate or Dithering (D)\cite{zhang2013characteristics} phase. In addition, when the plasma is in H mode, Edge Localized Modes (ELMs) can periodically occur. 

Current tokamaks regularly run in H mode, which motivates the necessity for some measure of control (and therefore, detection) of ELMs and transitions between plasma modes. Furthermore, it is expected that future machines will also run in the same operating conditions\cite{loarte2014progress}. Thus, the development of automated, data-based approaches to automatically detect the occurrence of certain events would be useful for both existing and future tokamak experiments and operation. A detector would not only simplify and speed-up the post-experimental, offline analysis of shots, but also (ideally) detect ELMs and plasma state rapidly enough to allow for its usage in the real-time control systems of a fusion experiment, for purposes of plasma control and real-time discharge monitoring and supervision\cite{humphreys2015novel}. 

Due to uncertainties in the scaling laws, it is difficult to determine, \textit{a priori}, when, during a discharge, a switch between different plasma modes will occur\cite{martin2008power}. Nevertheless, physicists can usually pinpoint, through a post-experimental visual analysis of several diagnostic signal time-traces, at what point in time any transitions between different modes did take place. Similarly to transitions between plasma modes, the occurrence of an ELM can usually be pinpointed by looking at the time-traces of several diagnostics from a plasma discharge post-shot. Yet through an analysis of signals, some types of ELMs can be easily confused with dithers; a distinction between the two phenomena can not always be clearly made\cite{ryter1994h}. 

Although the identification by an expert, through post experimental visual analysis of signal time-traces, of a single ELM, or a single transition between plasma modes, is relatively straightforward for a typical shot, it becomes much more cumbersome to carry out that analysis effectively for many shots, especially when the associated time-series data is long, and when a shot has many transitions between different modes.   

Recent advances in the ML field with the introduction of Deep Learning (DL) approaches deal with exactly such challenges. In the past years, the field of Deep Learning has brought about significant advances in Computer Vision and Sequential Data Processing. Convolutional Neural Networks (CNNs) have proven adept at localization, recognition and detection tasks in both 2-dimensional\cite{Szegedy_2015_CVPR, ince2016real, NIPS2012_4824, Tompson_2015_CVPR, lahivaara2018deep} and 1-dimensional\cite{acharya2018deep, abdeljaber2017real, malek2018one, golik2015convolutional, kiranyaz2015real, ronao2016human} data (i.e. signal analysis) in many different fields of science. In addition,  Long Short-Term Memory (LSTM) Networks, which are one type of Recurrent Neural Network, have been successfully used for processing of sequential data where one expects correlations to exist across time, namely, automatic translation, natural language modelling\cite{sundermeyer2012lstm}, traffic analysis\cite{ma2015long}, and automated video description\cite{venugopalan2016improving}. These tasks are much akin to what one can expect to find in terms of processing fusion shot data. 

Given this, a Deep Learning approach is well motivated to address this challenge. Specifically, deep neural network models offer particular advantages when modeling high-dimensional data as given in this setting. In this work, we develop an approach for automatic classification of L-D-H plasma states and detection of ELMs based on two deep neural network models. The first model is based on a sliding-window feed-forward neural network, specifically a convolutional neural network (CNN). The second model is based on a recurrent neural network (RNN), specifically a long short-term memory network (LSTM) with convolutional layers. 
The first model captures the local correlations within the windows to classify the transitions between plasma states from the shape of the signals. The second model extends this to capturing longer-term dependencies in the evolution of the states with the recurrent neural network layers. 

We empirically demonstrate the approach on data collected from the TCV tokamak experiment, labelled by an ensemble of experts. The presented results demonstrate the effectiveness of the proposed model to detect the state and events of the plasma. We further discuss the trade-offs between increased precision and increased complexity of both models.
 
This paper is organized as follows: Section \ref{sec:previous} discusses related work and Section \ref{sec:background} describes the physical phenomena being analyzed. Section \ref{sec:methods} formalizes our problem, details the data we have available, and explains our decisions regarding how we model the data and design and train the neural networks. Section \ref{sec:metrics} gives an overview of the metrics we used to evaluate our results and our rationale behind using those metrics. Section \ref{sec:results} gives an overview of the results achieved, and we wrap up with a discussion in Section \ref{sec:conclusions}.

\section{Previous work}\label{sec:previous}

Several different approaches for automated detection of events in plasma experiments exist. One such approach is to use threshold-based detectors. This corresponds to defining a point or series of points (in time) at which a signal surpasses a certain amplitude as corresponding to a detection\cite{webster2013statistical, greenhough2003probability, shabbir2014discrimination}, with additional constraints such as an increasing probability of the occurrence of an ELM as time passes since the last one. These approaches are limited to simple thresholding and cannot compute complex patterns in the data. Other work builds upon methods such as Kalman Filters to model the expected characteristics of the signal over a period of time\cite{shousha2019detector}, whilst also keeping track (in each time point) of the current plasma mode, according to a pre-defined model. In both of these cases, a detection algorithm's performance depends on the extent to which the theoretical assumptions and mathematical descriptions as to how the signals should behave are correct, whether those assumptions are exhaustive (i.e., whether there may be additional causes which are unaccounted for), and whether some of those assumptions are more important than others; in other words, it is difficult to design an exhaustive rule-based system to detect the occurrence of transitions between plasma modes, as well as to detect ELMs. 

The alternative is to use a purely data-based, supervised, Machine Learning (ML), approach, whereby a set of data, previously manually labeled by an expert (for example, through visual analysis), is used to train a detector. In this case, one does not specify which characteristics or correlations in the data are thought to correspond to the occurrence of an event; rather, it is expected that the algorithm can automatically learn what those correlations are, based on the labels, and then use the learned data features to make correct classifications on new data. Examples of such work are the usage of Support Vector Machines (SVMs)\cite{vega2009automated, gonzalez2012automatic, murari2006fuzzy, lukianitsa2008analyses} and Multi-Layer Perceptron (MLP) Neural Networks\cite{meakins2010application} on data from several tokamaks for detection of L-H transitions, classification of L and H modes, and detection of ELMs.  

This type of scenario is, indeed, well suited for application of ML methods towards enabling automation. However, traditional ML methods such as SVMs and MLPs typically have limitations when faced with data with complex dynamics, such as the long sequences (i.e., signal time-series) present in this environment.  SVMs typically depend on expert-defined feature engineering, which, while being superior to simple threshold-based detectors, is nevertheless insufficient when considering the complex data correlations which are observed in this setting. On the other hand, MLPs, while not requiring that sort of expert-defined input, are very inefficient when compared to modern Deep Learning models such as CNNs and RNNs, requiring much larger numbers of neurons and layers to perform the same task. These limitations are what motivate us to use Deep Learning approaches instead.

\section{Background} \label{sec:background}

\subsection{Low, dither and high plasma confinement modes}

When a discharge starts, the plasma is considered to be in Low (L) confinement mode. Once a certain threshold of input heating power to the plasma is reached\cite{xu2014dynamics}, the plasma can spontaneously transition into High (H) confinement mode. Originally discovered at the ASDEX-Upgrade Tokamak\cite{wagner1982regime}, High (H) mode is nowadays regularly observed in almost all other machines\cite{PhysRevLett.60.2276}. H mode is characterized by the appearance, in the plasma edge, of a steep gradient in the electron density and the electron/ion temperatures, and a reduction in the transport of particles and energy. As a consequence of this edge transport barrier, the temperature and energy in the plasma core increase. When compared to L-mode, H mode allows for a larger amount of stored plasma energy per input power, thus rendering the fusion process more efficient. Yet the actual input power threshold that triggers the transition between the two modes is dependent on many factors, such as, for example, the configuration of the magnetic field, plasma density, and plasma size \cite{martin2008power}. Furthermore, when the input heating power passes the aforementioned threshold but a change from L to H mode does not immediately occur, the plasma can be considered to be in a dithering (D)\cite{zhang2013characteristics} phase. In this case, a temporary, weak, edge transport barrier starts to develop at the plasma edge, only to collapse and reappear in rapid succession\cite{xu2014dynamics}. These oscillations then repeat themselves until the plasma transitions into L or H mode. The localization of transitions into, and out of, D mode can, however, be difficult to identify, and there are often disagreements between experts as to which periods of a shot are in a Dithering phase \cite{basse2003characterization}.

\subsection{Edge Localized Modes}

When the plasma enters H mode, the corresponding accumulation of energy and the large pressure gradient at the plasma edge can trigger the occurrence of Edge Localized Modes (ELMs). These consist of periodic bursts of particles and energy which, if a long amount of time passes between successive ELMs, can impose a significant power load on the divertor, potentially damaging it. However, ELMs also allow for the periodic removal of accumulated impurities from the plasma, and for a relaxation of the plasma density, which can otherwise increase as the H mode progresses, eventually triggering a disruption\cite{martin2001elming}. On the other hand,  frequent, less energetic, ELMs lower the power load on the divertor, at the cost of reduced plasma confinement. Thus, tokamak operation requires knowledge of the occurrence of ELMs, in particular for larger machines where ELMs may cause deterioration of in-vessel components. Although several different types of ELMs exist, for the purposes of this work, we did not make any distinctions between them -- we train the models to detect all occurring ELMs equally, regardless of their subclass.

\section{Methods} \label{sec:methods}

\subsection{Problem formulation and approach} \label{subsec:loc1d}
To develop a model for this task, we formulate the problem as follows:\\
We observe a sequence of measurements $x_t$ for $0 < t \leq N$ from the sensors for each shot. These observations are conditioned on the state of the plasma $z_t$ at corresponding time $t$, where $z_t \in Z$ and $Z: \{'Low', 'Dither', 'High'\}$. Our goal is to find the most likely sequence of $z_{1:N}$ and the occurrence of ELMS $e_{1:N}$ that explains the observations $x_{0:N}$. 

$$\hat{z}_{1:N} = \argmax_{z_{1:N}} \sum_t \log p(z_t|x_{0:t}, z_{t-1})$$
$$\hat{e}_{1:N} = \argmax_{e_{1:N}} \sum_t \log p(e_t|x_{0:t})$$

For this purpose, we develop two models. 

The first model is trained to detect the transitions between the different states of the plasma defined as $q_t \in Q$ where $Q:\{$  $'Low \rightarrow Dither'$, $'Dither \rightarrow Low'$, $'Low \rightarrow High'$, $'High \rightarrow Low'$, $'Dither \rightarrow High'$, $'High \rightarrow Dither', 'No transition\}$ and to detect the ELM events as $e_t \in E$ where $E: \{'ELM', 'No ELM'\}$. 

We implement this model with a feed-forward CNN that processes a window of observations $x_{t-w},.., x_{t}, ..., x_{t+w}$ and produces a probability distribution over the transitions $p(q_{z_{t-1} \rightarrow z_t}|x_{t-w:t+w})$ and over the presence of an ELM $p(ELM_t|x_{t-w:t+w})$ at $t$.

We now define the probability of transitioning to $z_t$ after being in $z_{t-1}$ ($p(z_t|x_{0:t}, z_{t-1})$) with our model $p(q_{z_{t-1} \rightarrow z_t}|x_{t-w:t+w})$ where $w$ is the number of observations around $t$, therefore:

$$\hat{z}_{1:N} = \argmax_{z_{1:N}} \sum_t \log p(q_{z_{t-1} \rightarrow z_t}|x_{t-w:t+w})$$

\begin{figure*}[h!]
    \centering
    \includegraphics[scale=0.4, trim=0 50 0 0, clip]{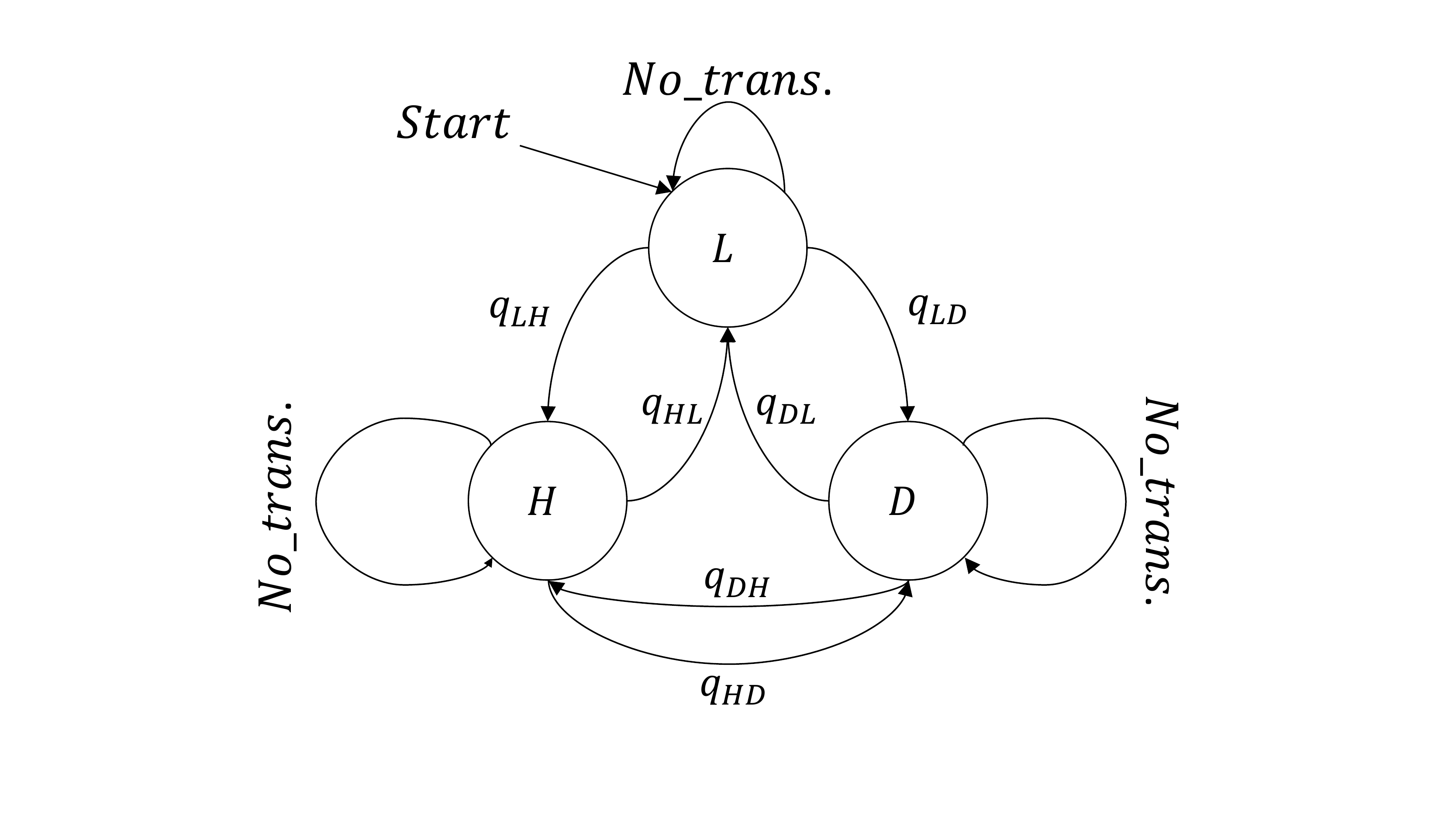}
    \caption{State machine for processing of the CNN outputs}
    \label{fig:state_machine}
\end{figure*}

Practically, we implement the $\argmax$ given above as a state evolution of a final state machine  $S_t(z^{(a)} \rightarrow z^{(b)})$ where $z^{(a)}$ and $z^{(b)}$ are elements in $Z$ and the transition probabilities are given by $p(q_{z_{t-1} \rightarrow z_t}|x_{t-w:t+w})$ at time $t$ (see Figure \ref{fig:state_machine}). The evolution of the state machine produces several possible sequences of states, and the one most likely to have generated the observed sequence of transitions can be found through an implementation of the Viterbi algorithm\cite{jurafsky2014speech}.

\begin{figure*}[h]
    \centering
    \includegraphics[scale=0.5, trim=0 180 80 0, clip]{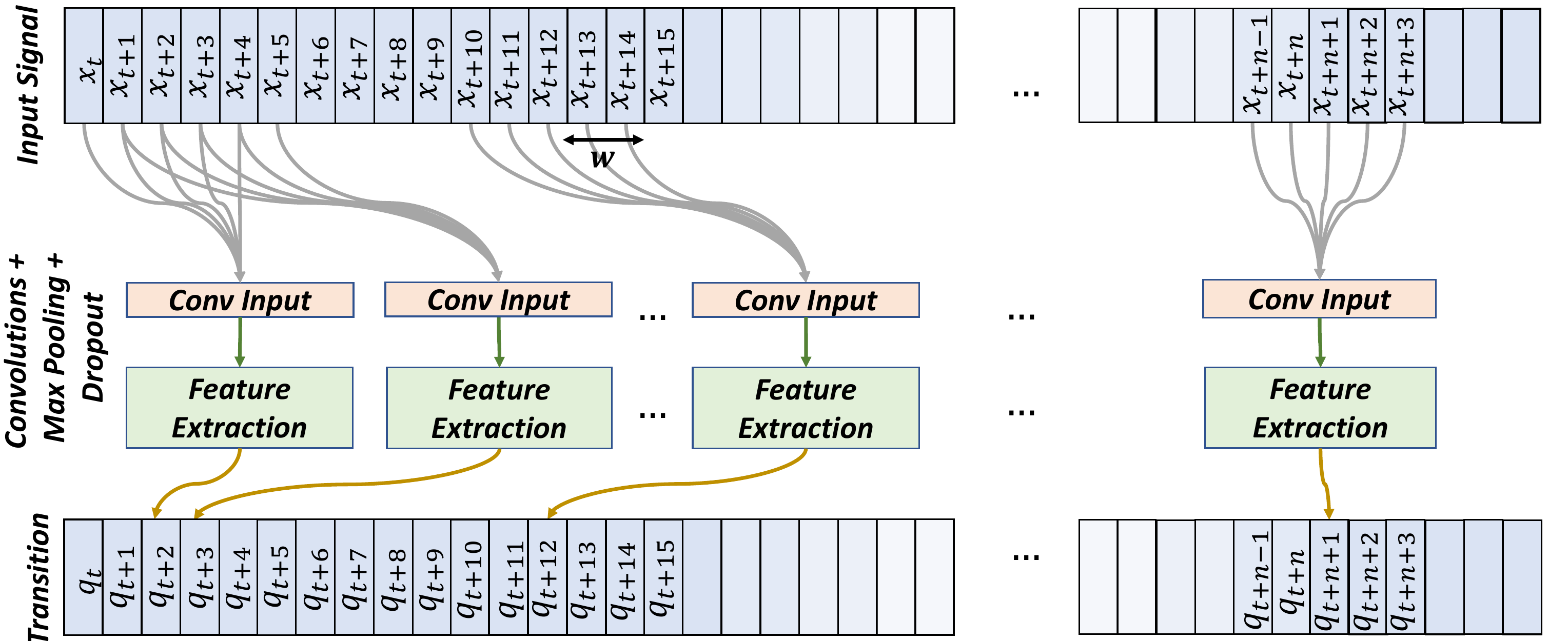}
    \caption{Representation of how a CNN can be used to model the transitions between different plasma modes. The network's output prediction for a time slice $t$ depends only on the data features in a defined region immediately surrounding $t$. }
    \label{fig:conv_dflow}
\end{figure*}

The first model can capture the localized correlations in the signals that indicate the transition of the state of plasma well. However, it is incapable of capturing the longer distance correlations that may be present in the signal. To generalize the approach further, we introduce a sequence model that models the full sequence of observations up to time $t$ and produce a probability distribution $p(z_t|x_{0:t})$ for $0 < t \leq N$, as well as a distribution over the presence of the $ELM$s ($p(ELM_t|x_{0:t}$). This model is implemented by extending the CNN with a recurrent (LSTM) neural network. In this case, the model now observes a sequence of sliding windows $x_{t-w}, ... , x_{t}, ..., x_{t+w}$ for each $t$ in the range $\{0, .. t\}$. 

\begin{figure*}[h]
    \centering
    \includegraphics[scale=0.5, trim=0 100 80 0, clip]{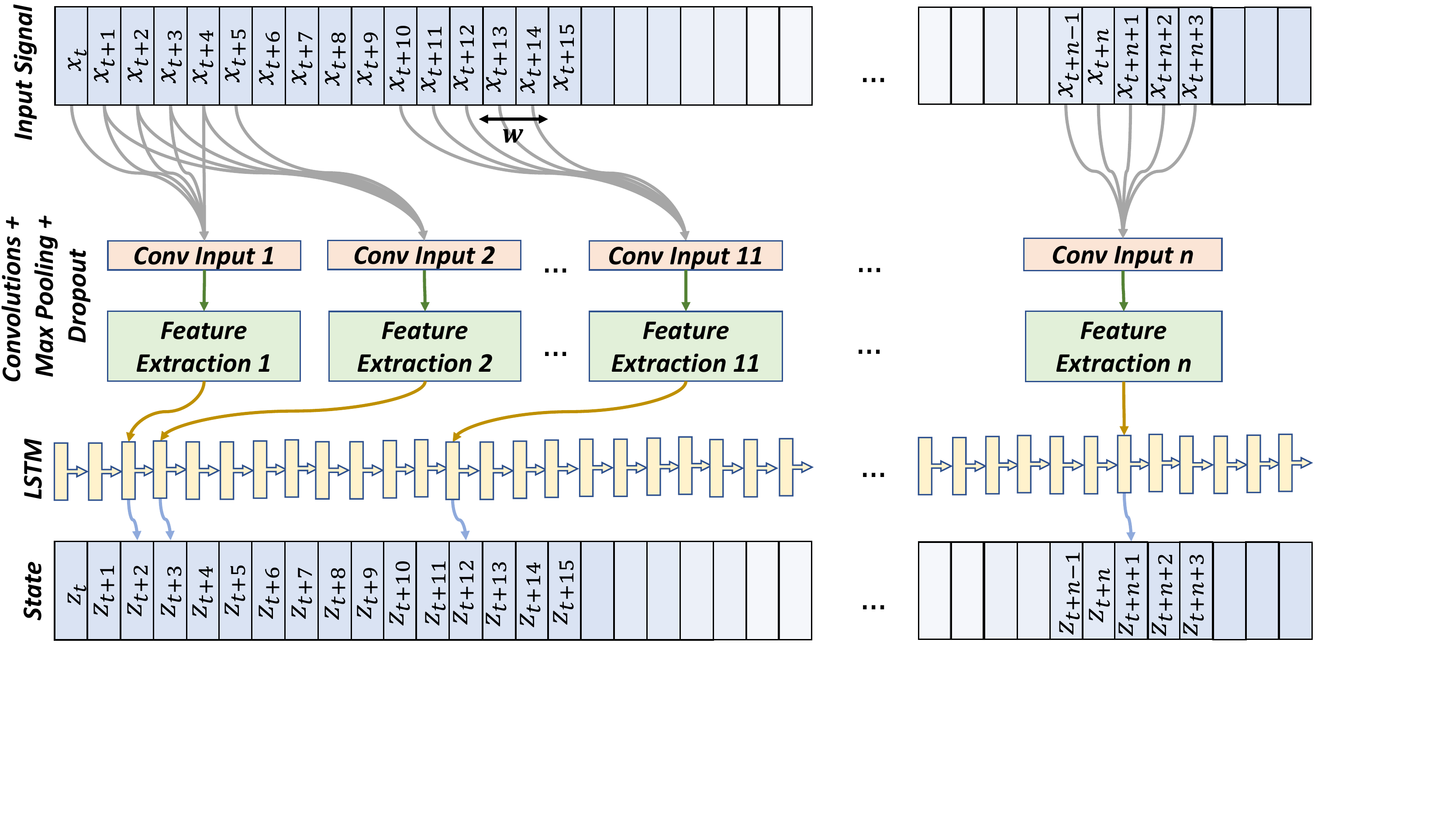}
    \caption{Schematic representation of the flow of data inside a convolutional LSTM Neural Network. The network's prediction (i.e. output probability) at any time $t$ of a shot depends not only on whatever features the convolutional layers have extracted from the points immediately around $t$, but also on features extracted in the past.}
    \label{fig:conv_lstm}
\end{figure*}

The first model has a lower computational complexity and can be trained more efficiently, as we only need windows of the signal with or without the different transitions, but it is limited to the information only present in the given window (see Figure \ref{fig:conv_dflow}). Increasing the size of this window that forms the context, increases the complexity both of the model and of dealing with multiple transitions appearing. 

The second model addresses these challenges by modeling the sequence rather than a fixed window (see Figure \ref{fig:conv_lstm}). As a sequential model, it has an internal representation of the past observations $x_0, .., x_t$, that enable it to weigh-in the likelihood of transition based on information in the more distant past\cite{boulanger2013high}. The LSTM effectively assumes the role of the finite state machine and so the model can directly model the state of the plasma rather than the transitions. However, it introduces higher level of complexity, particularly for training, as we need to train on sequences rather than fixed-length windows.  

\subsection{Data and event features} \label{subsec:datafeatures}

For the purposes of this work, we have assembled a dataset based on the time-traces of four signals originating in the TCV tokamak\cite{hofmann1994creation, coda2019physics}. We opted, for the purposes of this work, to use the same, limited set of diagnostic signals that experimentalists use to determine, in post-shot analysis, the state of the plasma (Figure \ref{fig:32195_full_shot_noelm}).

\begin{figure}[h]
    \centering
         \includegraphics[
         width=\textwidth,
         trim= 0 0 0 0,
         clip]
         {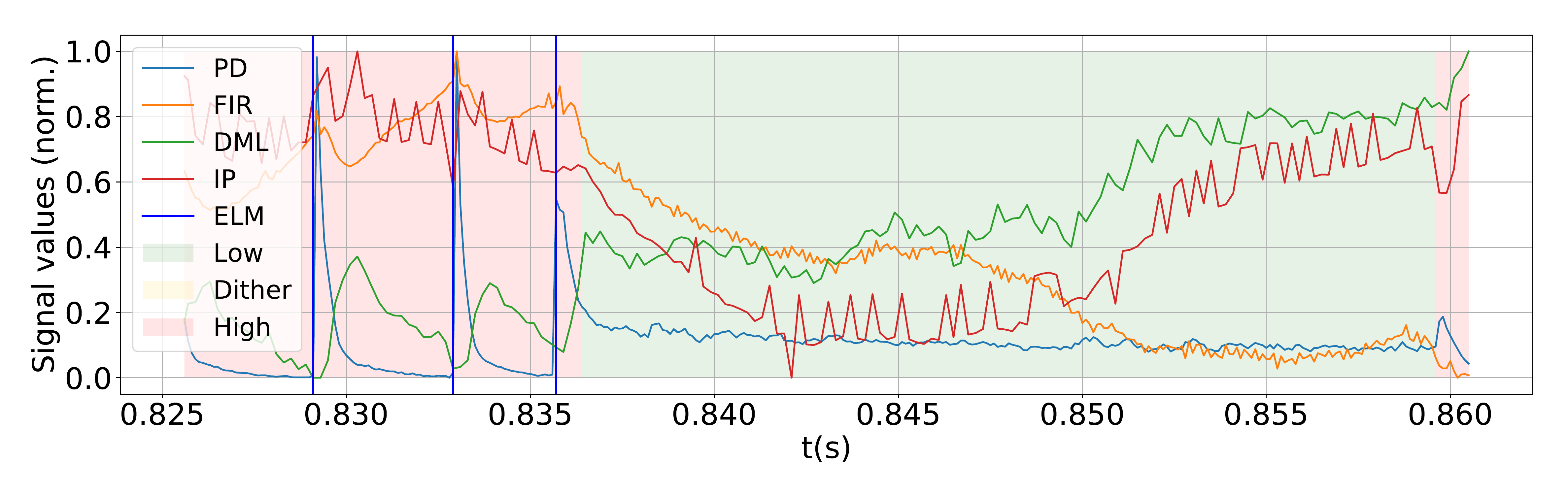}
         \caption{ELMs, and L and H modes from a section of TCV shot \#31650.}
        \label{fig:31650_elm_lh_hl_all}
\end{figure}

\begin{figure}[h]
    \centering
        \includegraphics[width=\textwidth,
        trim= 0 0 0 0, 
        clip]{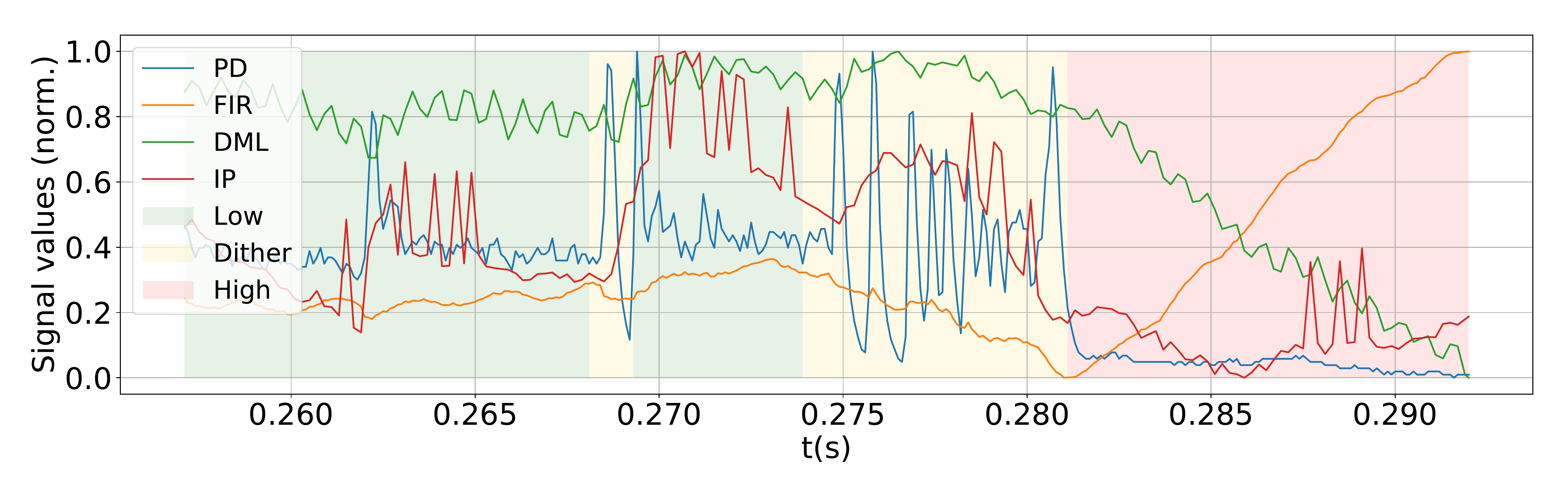}
        \caption{L, D and H modes from a section of TCV shot \#32195. }
        \label{fig:32195_dh_ld_dl_all}

\end{figure}

\begin{enumerate}

\begin{figure}[h]
    \centering
        \includegraphics[
        width=\textwidth,
        trim= 0 0 0 0,
        clip]
        {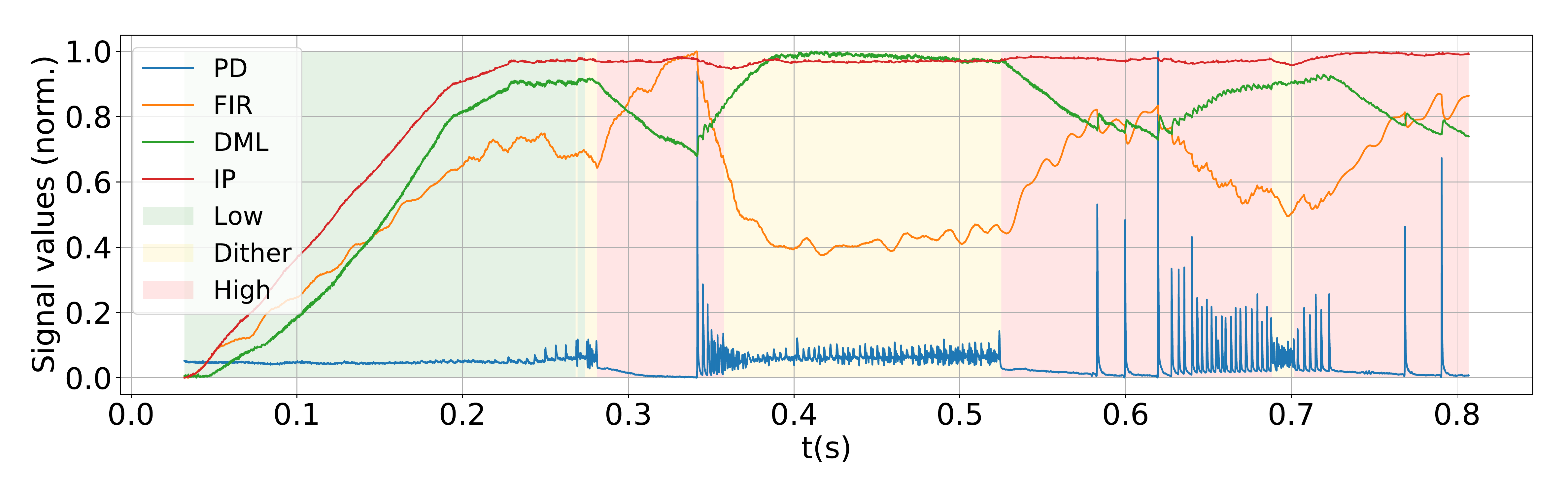}
        \caption{Switches between different plasma modes(Low, Dither and High), and time-traces of the collected signals, TCV shot \#32195}
        \label{fig:32195_full_shot_noelm}
\end{figure}%
    
\begin{figure}[h]
    \centering
        \includegraphics[width=\textwidth,
        trim= 0 0 0 0, 
        clip]{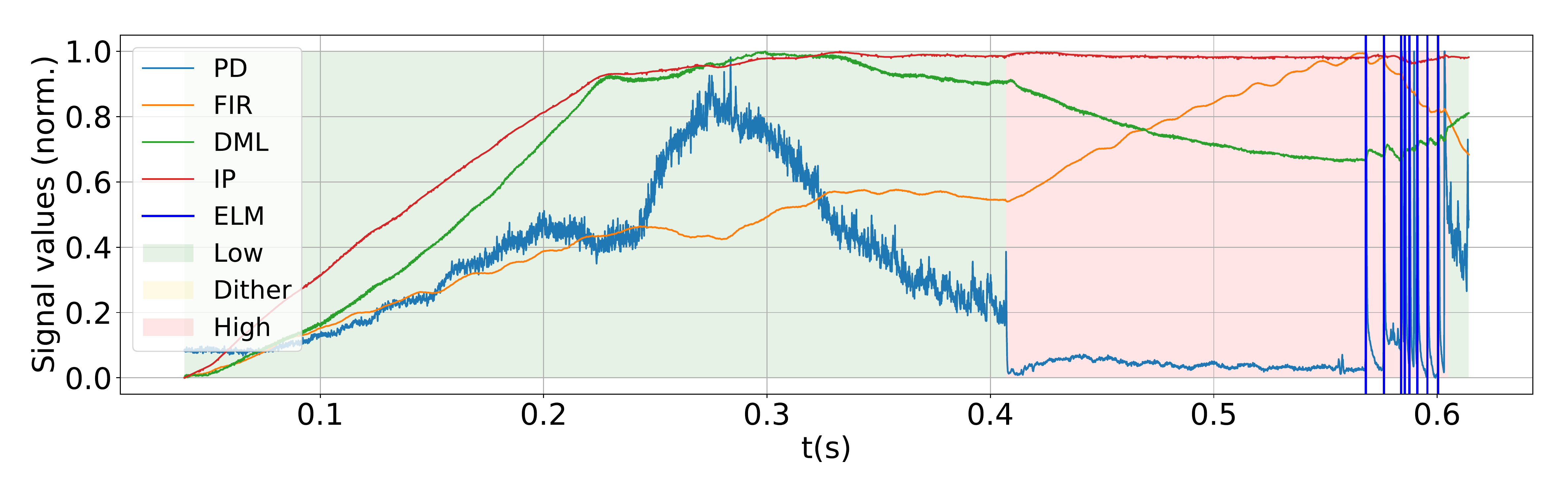}
        \caption{ELMs and L and H plasma modes, TCV shot \#33446}
        \label{fig:33446_full_shot}

\end{figure}

\item \textit{Photodiode (PD) signal}. Corresponds to the measurements given by the photodiode diagnostic at TCV along a vertical chord,  measuring the line-integrated emitted visible radiation; the photodiode has an $H_{\alpha}$ filter which measures radiation at 653.3 nm.

Transitions between different plasma states, as well as ELMs, can be most easily observed through analysis of the photodiode (PD) signal (Figure \ref{fig:33446_full_shot}). Transitions from L to H mode are characterized by a sudden drop in the baseline value of the signal, whereas transitions back into L mode have the opposite trace, i.e., the baseline PD signal suddenly increases and remains at a steady level. ELMs are characterized by a sudden spike in the PD signal, followed by a relaxation that takes at most 2ms. D modes generate rapid fluctuations in the signal (see Figure \ref{fig:32195_dh_ld_dl_all}); they do not necessarily correspond to a change in the baseline signal value, unless they are followed by a transition into a different state from the one at the point where they started.

\item \textit{Interferometer (FIR) signal}. The interferometers at TCV measure the line-integrated electron density in the plasma along 14 parallel, vertical lines of sight. Of these, we take the mean value, per time instant, of the 12 inner-most detectors.  

In the interferometer signal, the transition between L and H mode can most easily be seen as a sudden increase in the time derivative of the signal, while transitions back into L mode correspond to a decrease in the derivative. Similarly to what happens with the photodiode signal, ELMs may provoke short (albeit less pronounced) spikes in the FIR signal. 

\item \textit{Diamagnetic Loop (DML) signal}. Refers to the measurement of the total toroidal magnetic flux of the plasma\cite{moret2003fast}. The derivative of the DML signal frequently switches signs when a transition occurs between L and H mode, as well as when an ELM occurs (Figure \ref{fig:31650_elm_lh_hl_all}). Furthermore, the sign of this signal's derivative changes depending on the sign of the plasma current. 

\item \textit{Plasma Current (IP) signal}. Refers to the total plasma electric current. For this work, we use the current value to determine when the actual classification of plasma states should begin. Specifically, we ignore, for classification purposes, time points where the absolute value of the current is lower than 50 kA. 

\end{enumerate}

The 4 different signals used for this work have different sampling rates. As a first step, we resampled all of them to the same frequency of 10kHz. Since each TCV shot is usually up to 2 seconds long, this means that our shot signal data consists of time-series of up to about 20.000 time slices.  
 
We want to train a classifier to recognize features in the data which allow for detecting the occurrence of ELMs and transitions between different plasma modes -- i.e., a supervised learning task. As such, the first step was to collect labels for each shot time series, through visual analysis taking into account the data features described above. The collected data was visually labeled by 3 different experts for the same shots. This means for some shots, the same regions might have different labels (namely, the experts might disagree on whether a certain part of a shot is dithering). Training the network with labels which are different in some regions has several potential advantages. For example, it compensates for any possible discrepancies in labeling originating from human error. It also allows us to incorporate the uncertainty in the labels into the network training process itself, that is, it acts as a form of regularization: if there is no majority agreement between experts regarding a section of a shot, then it is to be expected that the network will also learn not to strongly favor any class in that region. Conversely, if the three experts agree, then the network will learn that the features in that region most certainly correspond to a certain class, which renders the classification more robust. Finally, getting labels from different experts allows us to increase the size of our training dataset.

\subsection{Model training} \label{subsec:data_encod_prep}

\begin{figure*}[h]
    \centering
    \includegraphics[scale=0.37, trim=100 0 100 50, clip]{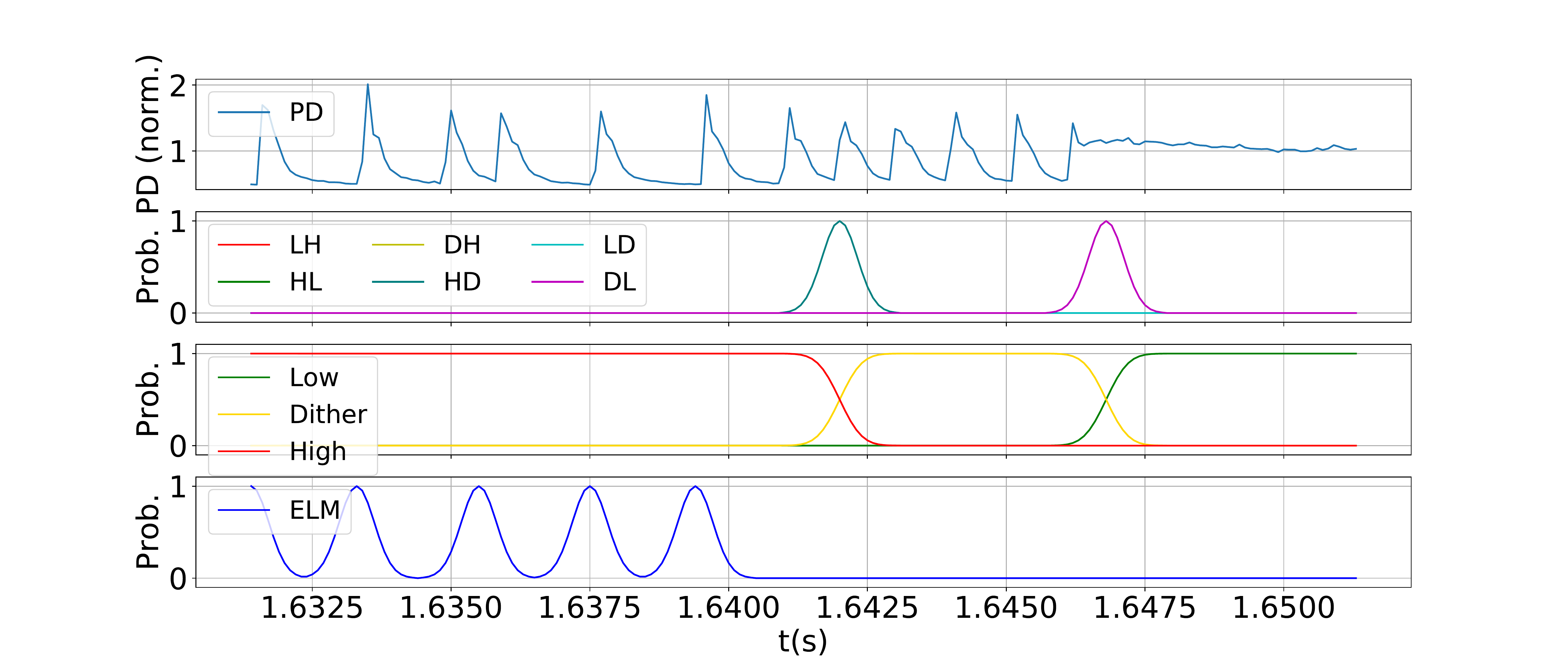}
    \caption{Representation of the different types of encoding of the target ``smooth'' data distributions, to be learned by the two classifiers, from TCV shot $\#30262$. Here, we show only the labels produced by a single expert, though the networks are trained with labels from all of them. The second plot from the top illustrates the transitions to be learned by the CNN, while the bottom-most plot illustrates the states to be learned by the Conv-LSTM.}
    \label{fig:data_encodings}
\end{figure*}

The two proposed models develop different maps. The first model is a map between a fixed window of observations and a distribution over transitions, while the second models a sequence of observations and produces a sequence of states (see Figure \ref{fig:data_encodings}).

\begin{figure}[h]
    \centering
    \includegraphics[scale=0.45, trim=0 30 0 50, clip]{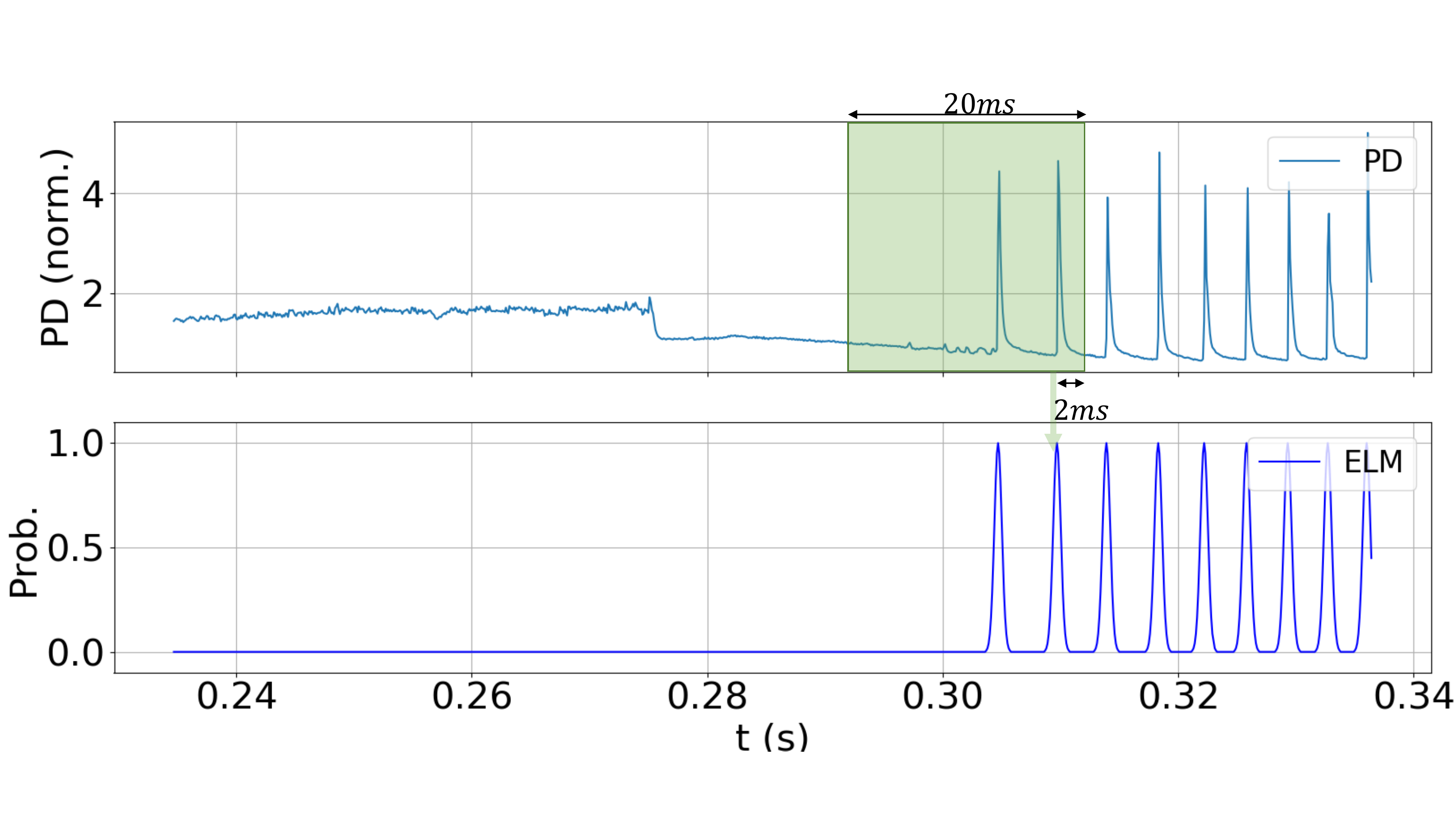}
    \caption{Representation of the sliding temporal windows fed to the CNN on top of the PD signal, and their corresponding ELM probability output. At inference time, these windows slide over the 4 signals across the whole shot, each of them rendering an output probability for a single time slice. }
    \label{fig:elm_conv_window_cnn}
\end{figure}

Accordingly, the training data has different arrangements. For transition classification, we need to prepare a dataset $D_1,\{(x, q, e)\}$, 
where a training point consists of a section of the recorded signal($x_{t-w}, ... x_t, ...,  x_{t+w}$), the corresponding label of one of the transitions $q_t$ in $Q$ and the matching label $e_t$ indicating the presence (or not) of an ELM. Figure \ref{fig:elm_conv_window_cnn} illustrates this in detail.

For the second model $D_2,\{(x, z, e)\}$, a training point consists of a sequence of windows of observations drawn from $x_t$ to $x_{t+l+w}$ (where $l$ is a defined sequence length, and $w$ is the window length), a sequence of state labels $z_t$ in $Z$ of length $l$, with each label corresponding to the state of the plasma at times $t$, and a sequence of labels $e_t$ of length $l$ corresponding to the presence of an ELM at times $t$. Figure \ref{fig:elm_conv_window_cnn} illustrates this in detail.

There is an inherent uncertainty in the labeling of the ELMs and plasma states, particularly when it comes to transitions into and out of dithers. The raw data only has hard, binary, one-hot encodings\cite{harris2010digital} -- that is, a transition between two states, for example, is labeled as a sudden switch (from one time slice to the next) from one state to another. This means that it is easy to mistakenly label an event or transition in a slightly shifted time slice. This type of hard threshold also makes it difficult for a neural network to generalize to outside of its training set\cite{szegedy2016rethinking}. 

Therefore, for the first model (CNN), we process the target time-series such that the probability of an ELM, or of a given state transition, is a continuous value, starting at zero and peaking at one, with several intermediate probabilities. In practical terms, we apply on each event a gaussian smoothing such that, if an ELM or state transition occurs at time $t$, its probability at that point is 1, and we define an interval $\Delta t$ -- before and after $t$ -- where the probability, respectively, smoothly increases and decreases. We defined these smoothing intervals as corresponding to 2ms, which, at the defined sampling rate, translates to 20 time slices. We do the same with the states $z_t$ for the second model (Conv-LSTM), such that a switch between two different states, from $z_1$ to $z_2$, does not happen immediately from one time slice to the next, but rather, the probability of $z_1$ decreases, while that of $z_2$ increases, over a span of 20 time slices. 

This procedure not only models the uncertainty in the labeling process, but also acts as an automatic regularization for the neural network training process, i.e., it makes it easier for a neural network to generalize what it learns to unseen data\cite{zheng2018improvement}.  

The choice of the size of the temporal windows with which the CNN is trained is a trade-off between the assumptions made about the data, and computational feasibility. Larger windows contain more spatial information and thus, intuitively, should make the classification at a particular time slice more precise, but also make the training and inference process by the network slower. Smaller windows contain arguably less information, but can be processed faster. We opted to train the CNN with temporal windows with a length of $20$ms, which we judged to be a good compromise between those two requirements. At our sampling rate, these windows are 200 time slices long. This is illustrated in Figure \ref{fig:elm_conv_window_cnn}: the green region represents a window of signals (in this case, only the PD signal) which is fed to the neural network, and its associated target, which is the probability of an ELM occurring at $t=0.304$s. There is an offset between the time at the window's rightmost edge, and the time for which the probability is computed; in the example of \ref{fig:elm_conv_window_cnn}, the offset is of $2$ms, which means that to detect the ELM occurring at $t=0.304$s, the window would have information on the signals from $t=0.286$s to $t=0.306$s. In formal terms, the windows compute in that case $p(e_t|x_{t-w_1:t+w_2})$ and $p(q_{z_{t-1} \rightarrow z_t}|x_{t-w_1:t+w_2})$, where $w_1 = 180$ and $w_2 = 20$. 
In practice, in a real-time setting, that offset would constitute a minimum delay between the occurrence of an event in a machine, and a detection by the classifier. Once again, the size of this offset is a trade-off: a smaller offset is ideal for real-time applications because it gives more time for feedback control mechanisms, but it also contains less information for the network to accurately classify an event. 

\begin{figure}[h]
    \centering
    \includegraphics[scale=0.45, trim=0 30 0 40, clip]{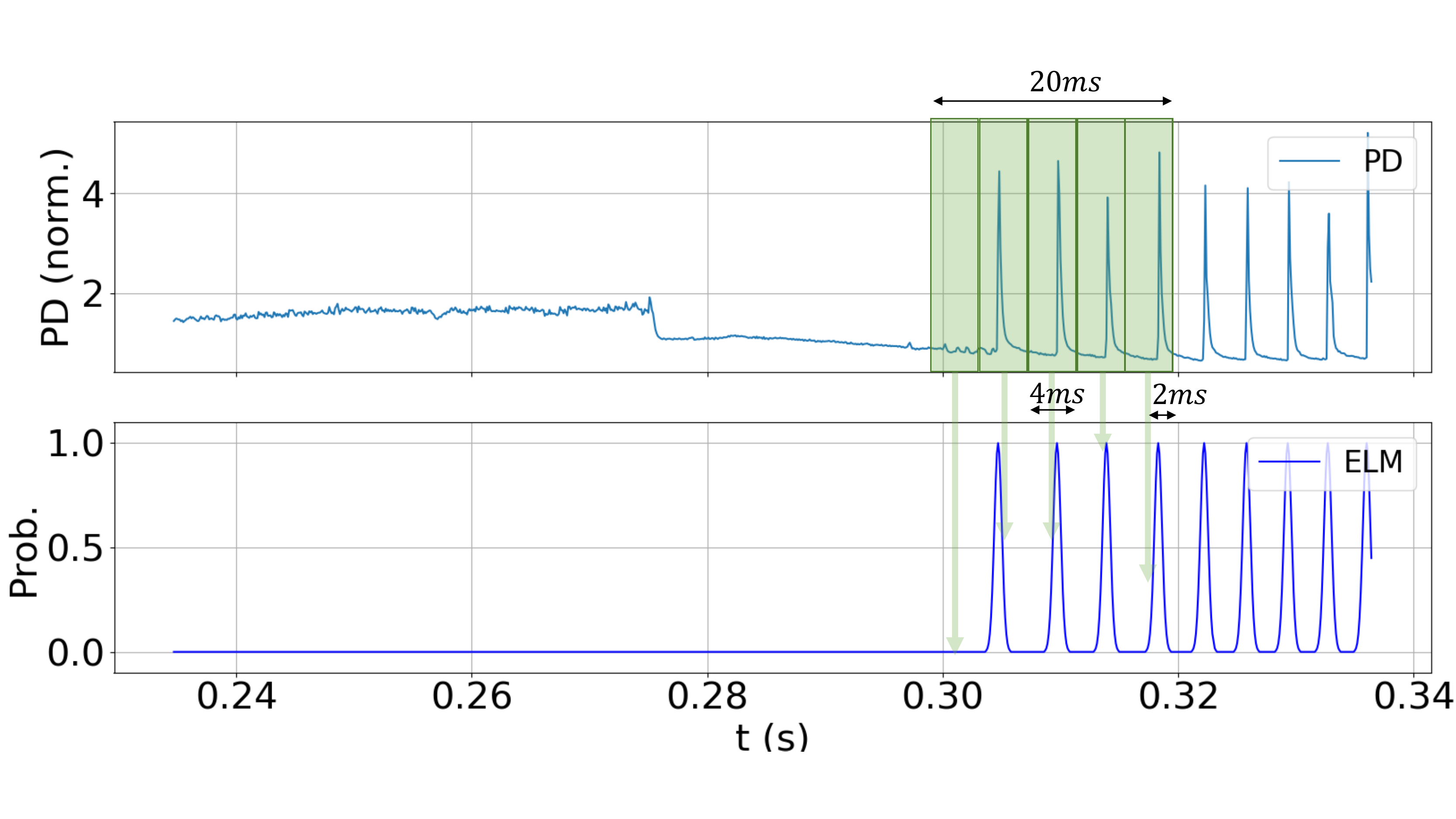}
    \caption{Example of a sequence fed to the LSTM. At a $10kHz$ sampling rate, it consists of 200 overlapping temporal windows of length 40. The output probability for a given window depends not only on what data features are present in that window, but also on the past windows in the sequence.}
    \label{fig:elm_conv_window_lstm}
\end{figure}

We train the Conv-LSTM not with windows, but with sequences of windows. The distinction is an important one, for it implies different assumptions about the data. In the case of the windows fed to the CNN, it is assumed that each window is independent of each other. In the data fed to the Conv-LSTM, each sequence itself is composed of several windows, with future windows depending on past ones. We defined each of those sequences to consist of 200 windows (since that was also the length of the windows fed to the CNN). In this case, each of the individual windows has a length of 4ms (40 time slices), with an offset of 2ms, as in the data for the CNN (see Figure \ref{fig:elm_conv_window_lstm}). The sequences have a stride\cite{dumoulin2018guide} of 1: each window starts and ends exactly 1 time slice after the previous one finishes. Each of these sequences is randomly subsampled from the whole shots, and the corresponding targets for them are chosen randomly from one of the three labelers. 

Although not all of these subsamples start in L mode, our expectation is that the network would learn by itself that an actual shot always begins in that state. There are several reasons for this. First, the network will learn to recognize any features in the subsequences that are consistent with the beginning of a shot, and learn that those features correlate to L mode. Second, even if some training sequences start in D or H mode, the network will statistically learn that these modes are more frequently the result of a transition from a previous mode.

\subsection{Model design}

\begin{figure*}[h]
    \centering
    \includegraphics[scale=0.6, trim=150 250 150 0, clip]{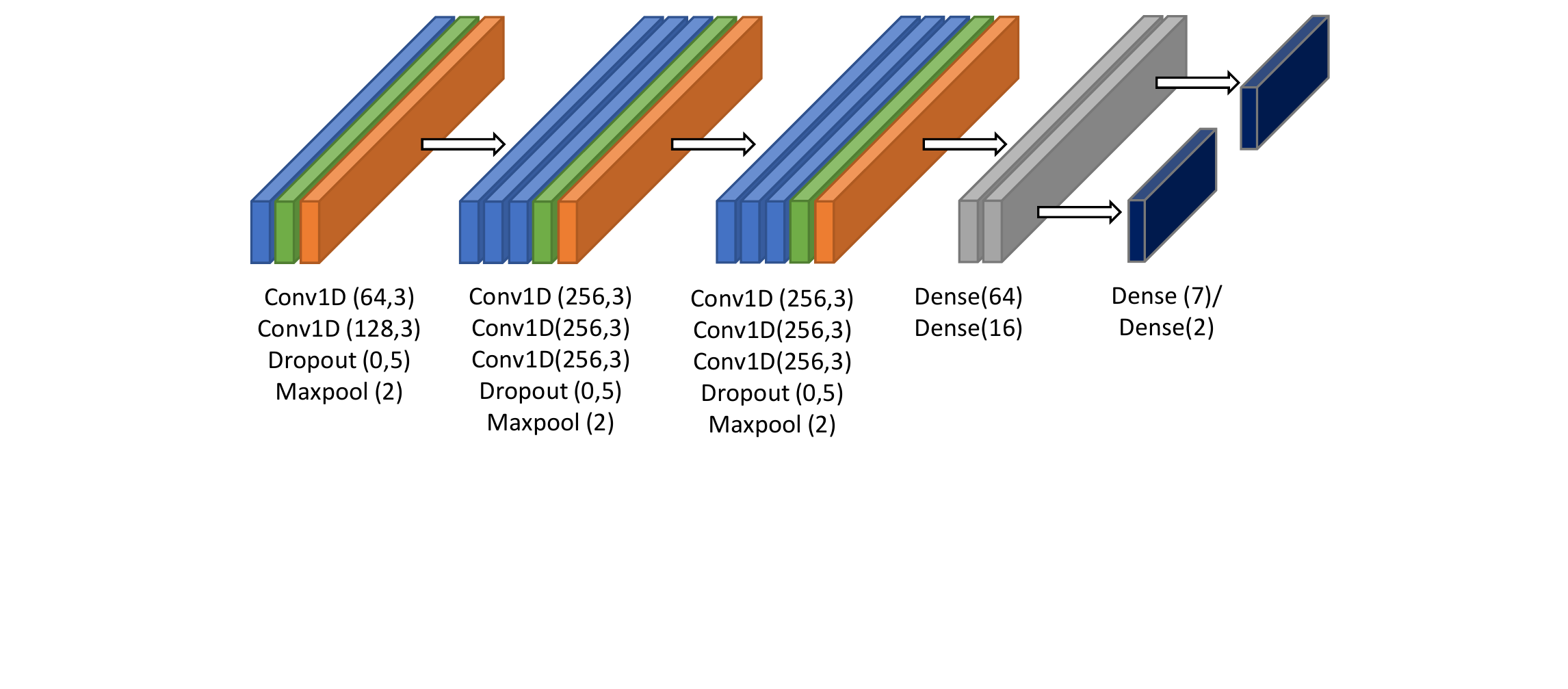}
    \caption{Architecture of the Convolutional NN }
    \label{fig:cnn_arch}
\end{figure*}

The architecture of the neural networks used for the transition detection starts with a 1-D convolutions with four channels, each of which receives the values from the PD, FIR, IP and DML signals. These are followed by several convolutional layers, interspersed with pooling and dropout layers, which are trained for feature extraction, with deeper layers extracting higher-level data features (Figure \ref{fig:cnn_arch}). The last layers of the network are fully-connected, and are responsible for receiving the pre-processed high-level features and producing an appropriate output for them, i.e., the desired classification. This model is loosely inspired by the VGG architecture for classification of images where fixed sized filters are used\cite{simonyan2014very}.

\begin{figure*}[h]
    \centering
    \includegraphics[scale=0.6, trim=150 300 150 0, clip]{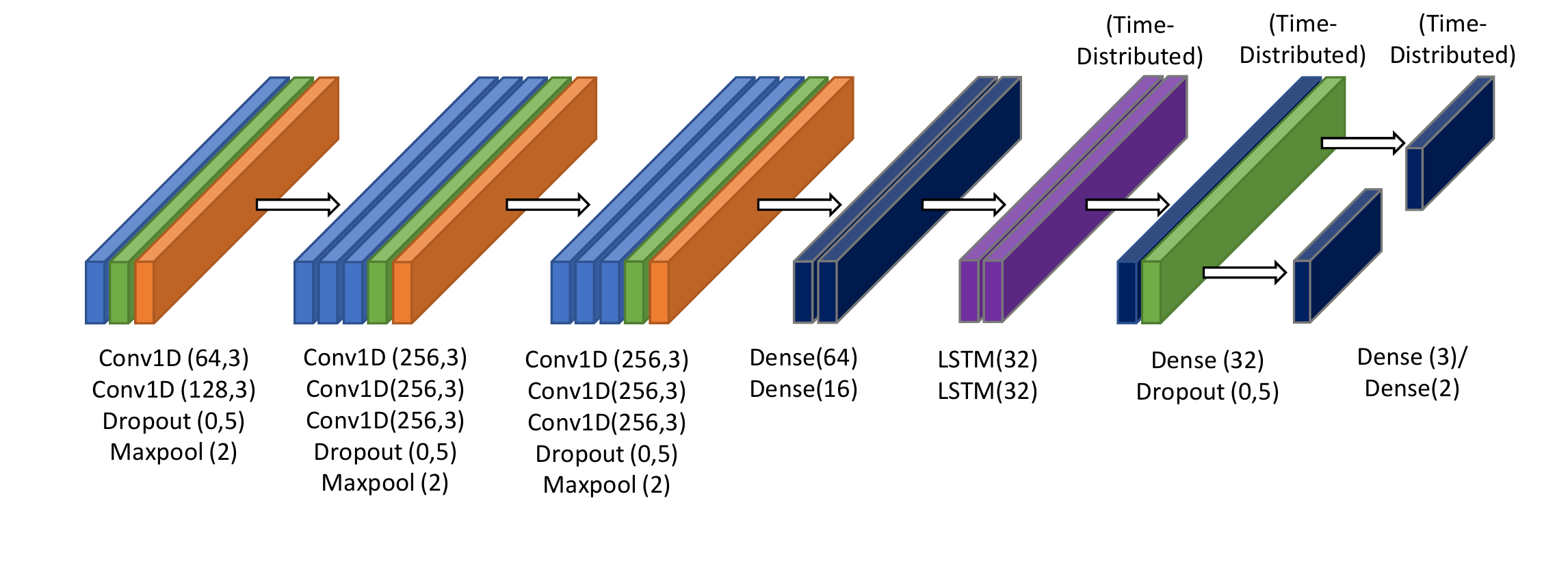}
    \caption{Architecture of the convolutional LSTM. All layers and nodes use ReLU activation functions, apart from the final output layer, which uses Softmax activation. }
    \label{fig:conv_lstm_arch}
\end{figure*}

Our convolutional LSTM network builds on top of CNN model that showed the best performance on the transition detection task. We add a recurrent layer that processes the output of the CNN to capture the longer-distance correlations in the data (Figure \ref{fig:conv_lstm_arch}). 

We designed the networks using the Keras framework for Deep Learning\cite{chollet2015keras}. Both networks used a categorical cross-entropy loss function, and were trained with the Adam optimizer\cite{kingma2014adam} using the default learning rate value provided by Keras.

\subsection{Data split}

In total, we possessed 54 shots fully labeled by the three experts. In a typical Deep Learning setting, some sort of normalization\cite{han2011data} is usually applied on the available data. The most common procedure would have been to normalize across the entire dataset. However, because of the different calibrations of the PD signals and the subsequent large variance and multimodal distribution associated with it, we decided, at this stage, to normalize each shot separately dividing each signal in each shot by its own mean across the whole shot. For potential real-time applications, as any new shots could fall outside the normalization range, the procedure would require grouping and normalizing the shots with respect to different signal gains and calibrations.

From these normalized full sequences, we draw batches of smaller temporal windows and subsequences to train the neural networks. There are several reasons for this subsampling.
First, the full shot time-series are up to about 20,000 time slices long, but the actual length of a shot can vary significantly. Yet for purposes of training the networks, we require batches of data of fixed length, which can be achieved by subsampling from the full sequences. 

Second, this method allows us to automatically perform data augmentation for training, since one long sequence will contain many shorter subsequences and windows. 

Third, feeding very large temporal windows to a CNN would be computationally difficult, as the number of network parameters requiring training would grow considerably. 

Finally, the distribution of the data in the full sequences is highly unbalanced: in most shots, dithering phases are significantly shorter than L and H phases; only a few dozen transitions happen at most per shot; and, some transitions tend to be more frequent than others. Training with whole sequences would significantly bias the networks towards the events and transitions that occur more frequently in the labeled data. Drawing subsequences allows us to control the data fed to the network such that this inherent bias is mitigated. To do this, the training data batches must be balanced, i.e., generated such that they contain roughly equal fractions of the different types of events and/or transitions of interest. In the CNN, there are 8 possible events of interest -- LH, HL, HD, DH, LD, DL, ELM, and no transition. Generating batches for the CNN means that, for a batch containing $n$ data samples, n/8 of those samples will correspond to each of those different types of transitions. Similarly, for the Conv-LSTM, the batches are generated such that the three target distributions (L, D and H) correspond to approximately 1/3 of the data samples each.

\section{Evaluation metrics} \label{sec:metrics}
\subsection{ROC curve}

We consider the detection of single, discrete ELMs by the networks as corresponding to a point in time (in a shot) where the direct network outputs for ELM probability $\hat{e}_{1:N}$ reach a maximum value. This is not necessarily a point where the output network probability for ELM is 1, but rather, a point $t$ where the output probability $P(ELM_t)$ follows a series of strictly increasing probability values, and precedes a series of strictly decreasing ones. Because we defined the length of the gaussian smoothing of the probabilities as $20$, here we consider a local maximum for $P(ELM_t)$ within a $20$-wide interval to correspond to the detection of a single ELM -- which we denote as a positive. The remaining points are considered non-detections, i.e., negatives. In addition, we defined different probability thresholds for what can be considered a detection of an ELM by the network. For example, defining a threshold of 50\% implies that only ELM probability maxima above that threshold are considered positives. 

Positives and negatives must then be compared to the labeled ELMs. To that end, we build the ELM Confusion Matrix, which defines several variables:  negatives that match their label at the same point in time are True Negatives (TN), while those that do not are False Negatives (FN). Similarly, positives that match their label are True Positives (TP) and those that do not are False Positives(FP).

Using this method to determine the points in which the network detects individual ELMs, one can then compute the True Positive Rate (TPR) and False Positive Rate (FPR) for different detection thresholds:

\begin{center}
    \begin{equation}
        TPR = \frac{TP}{TP + FN}
    \end{equation}
\end{center}{}

\begin{equation}
    FPR = \frac{FP}{FP + TN}
\end{equation}

Plotting the TPR versus FPR for a series of different detection thresholds yields the classifier's ROC curve\cite{fawcett2006introduction}, which illustrates the network's capacity for discrimination given different detection thresholds. There are several ways to compute the ideal detection threshold based on the ROC curve, depending on the task in question. In our case, we use the Youden index\cite{liu2012classification}, whereby the best threshold is the value which maximizes the difference $TPR - FPR$, the maximum value being $1$.

\subsection{Kappa statistic}\label{subsec:kstat}

To compare the models' accuracy with that of the human labelers, we use Cohen's Kappa-statistic coefficient, which measures agreement between two sets of categorical data\cite{landis1977measurement}, defined as 

\begin{equation}
    \kappa = \frac{p_0 - p_e}{1 - p_e}
\end{equation}

where $p_0$ denotes the actual relative agreement between the two sets, and $p_e$ denotes the probability of the two sets randomly agreeing with each other. Generically, the $\kappa$ coefficient's values oscillate between 0 and 1, the former indicating poor performance, and the latter indicating perfect performance. In our case, given two sequences $z_1$ and $z_2$ of plasma states, Kohen's Kappa measures the overlap between them. If $z_{1_t} = z_{2_t}$ for all time instants $t$, the metric will yield a score of 1; if there are mismatches between the two sequences, the score will go down.

The $\kappa$-statistic can be interpreted differently based on the sections of the data for which it is computed. For that reason, we will now define several variables that allow us to interpret the $\kappa$-statistic scores. 

Remember that we possess labels drawn from three different experts; as such, generically, labeled shot states at each point in time $t$ of a shot can be in one of three possible categories:
\begin{itemize}
    \item No majority agreement, i.e., all labelers disagree as to what state the plasma is in, which we denote as category $C_1$.
    \item  Majority agreement, i.e., two labelers agree on the state of the plasma, while one disagrees, which we denote as category $C_2$.
    \item  Consensual agreement -- all labelers agree as to what state the plasma is in, which we denote as category $C_3$. 
\end{itemize}{}

We define the union of $C_2$ and $C_3$ as \textit{ground truth} ($C_4$), i.e., they are sections of shots where there is at least a majority opinion as to what state the plasma is in. We also have, for each shot, the most likely sequences $\hat{z}_{1:N}$ of states (given the observed data) produced by the neural networks, which we will now denote as $C_5$. 

Computing the $\kappa$-statistic score, $\kappa_l$, between sets $C_2$ and $C_4$ gives an indication of the probability that a single labeler disagrees with the ground truth: a $\kappa_l$-score of 1 would indicate that there is agreement between all the labelers all the time, while a lower score would indicate that at least some of the time, one labeler disagrees with the others. Simultaneously, computing the $\kappa$-statistic score between sets $C_5$ and $C_4$ ($\kappa_n$) gives an indication of the networks' performance given the ground truth. But, in addition, we can directly compare $\kappa_l$ and $\kappa_n$. This comparison allows to test how a network and a single labeler compare against each other, on average, given the ground truth. 

The $\kappa$-coefficient is calculated separately for each of the three possible labels for the plasma state (L, D and H), and as a weighted mean across all three states. The weights of that mean are taken to be the relative frequencies of each individual state in the dataset, based on the ground truth ($C_4$) labels.

\section{Results} \label{sec:results}

We performed several training runs using the data labeled by the three experts; we carried out experiments where we trained both models (CNN and Conv-LSTM) three times, each time randomizing the training and test shots, to test whether differences in the data could lead to different results. In a typical Deep Learning setting, the data is usually split so that approximately $80-90\%$ is used for training, and $20-10\%$ is used for validation of the results, i.e., testing the network's capability to accurately predict on data that was not used for training. In our case, we opted for a training/test data split of $50\%$, i.e., of the 54 shots, we used 27 for training and 27 for testing. The results that follow are the best results of those three experiments, for each model. We also experimented with varying offsets (see Figure \ref{fig:elm_conv_window_cnn}) for the convolutional windows to see what effect that factor could have on the results; we settled for an offset value of $2ms$ (20 time slices), as smaller offsets degraded results, while larger ones did not improve them. We computed the metric scores on the training and test data at several points during training to control for overfitting\cite{bishop2006pattern}, and present the results from the epoch where the state detection results on test data were the highest. We ran the neural networks on an NVIDIA Quadro RTX 5000 GPU. 

\subsection{CNN}

We computed the $\kappa$-statistic based on the regions defined in Subsection \ref{subsec:kstat} -- that is, we compute scores based on the network output versus the ground truth ($\kappa_n$), and based on labeler disagreement versus the ground truth ($\kappa_l$). We computed the scores on a per-state (L, D and H) basis, and also computed a mean of the values obtained for each state.

\begin{figure}[h!]
    \centering
    \begin{subfigure}{0.3\textwidth}
        \centering
        
        \includegraphics[width=\textwidth,
        scale = 0.6, 
        trim= 0 0 0 0,
        clip]{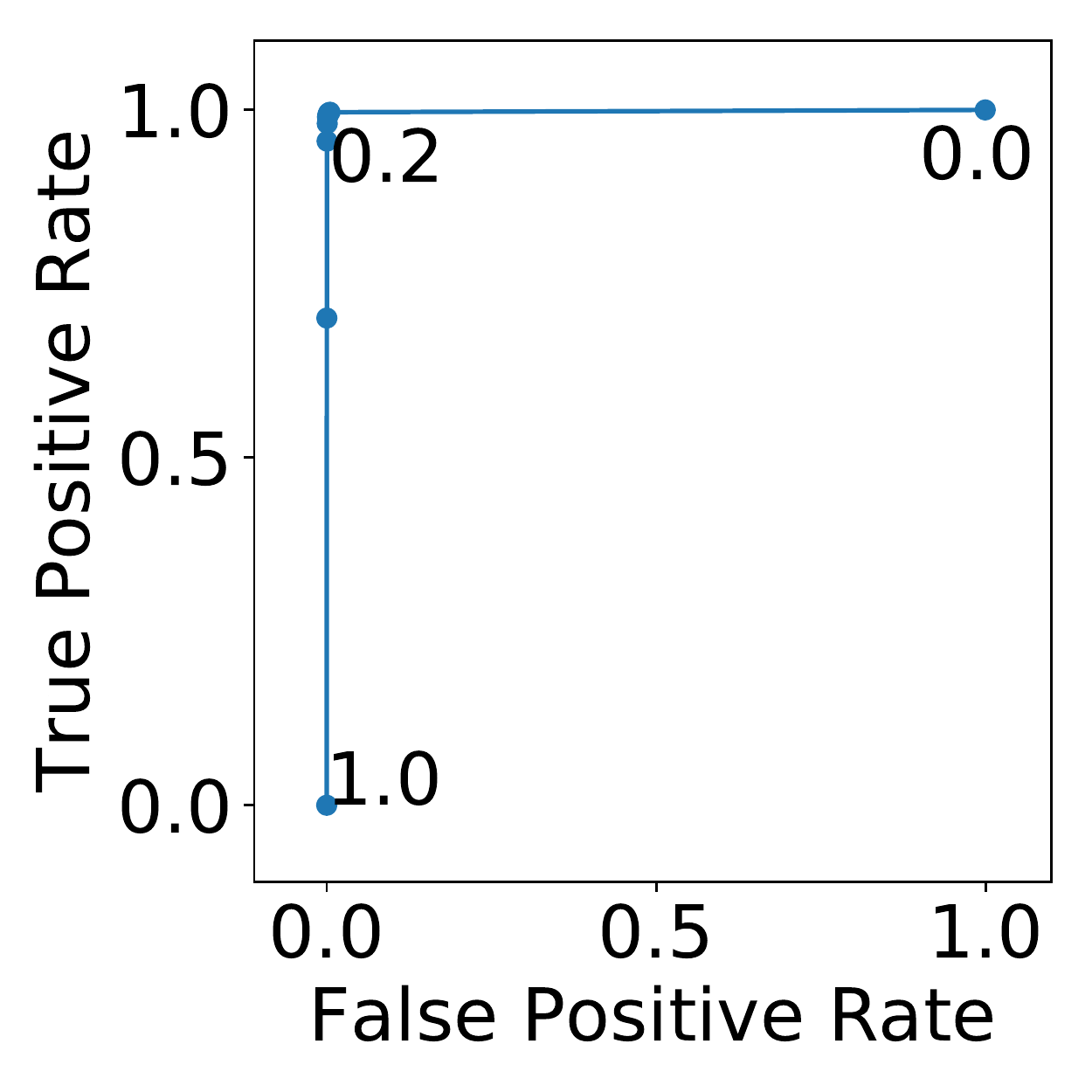}
        \caption{Training data.}
        \label{subfig:cnn_ROC_train}
    \end{subfigure}%
    ~
    \begin{subfigure}{0.3\textwidth}
        \centering
        
        \includegraphics[width=\textwidth,
        scale = 0.6, 
        trim= 0 0 0 0,
        clip]{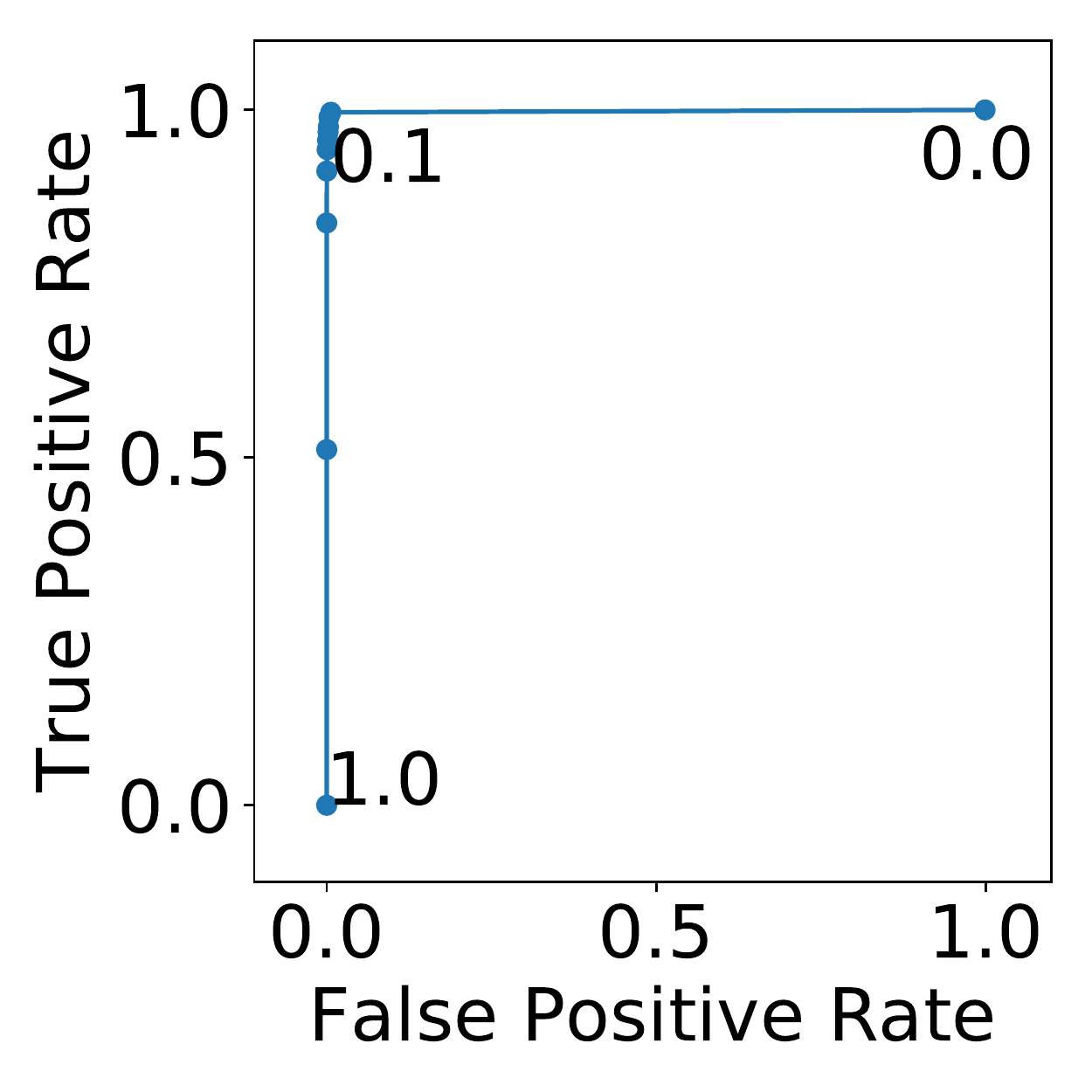}
        \caption{Test data.}
        \label{subfig:cnn_ROC_test}
    \end{subfigure}%
    \vspace{-3mm}
    \caption{ROC curves for ELM detection for the CNN model. The detection thresholds that maximize the Youden index are $0.2$ and $0.1$ for training and test data, respectively yielding index values of $0.993$ and $0.99$. Using the ideal threshold for the training data ($0.2$) on the test data gives a slightly lower Youden index of $0.986$.}
    \label{fig:cnn_roc}
\end{figure}

\begin{figure}[h!]
    \centering
    \begin{subfigure}{0.45\textwidth}
        \centering
        
        \includegraphics[width=\textwidth,
        scale = 0.55, 
        trim= 0 0 0 0,
        clip]{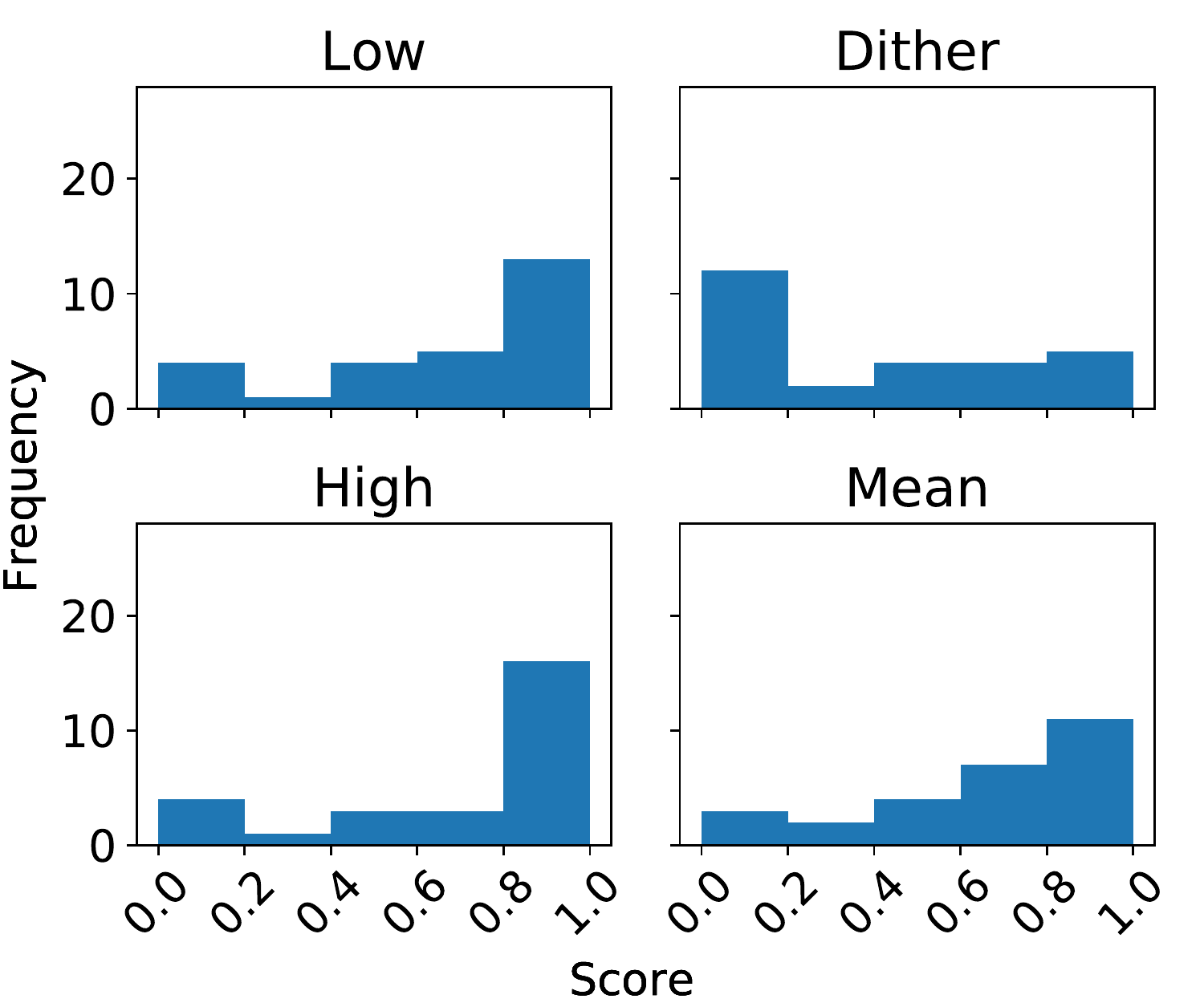}
        \caption{Training data.}
        \label{subfig:cnn_train_histo}
    \end{subfigure}%
    ~
    \begin{subfigure}{0.45\textwidth}
        \centering
        
        \includegraphics[width=\textwidth,
        scale = 0.55, 
        trim= 0 0 0 0,
        clip]{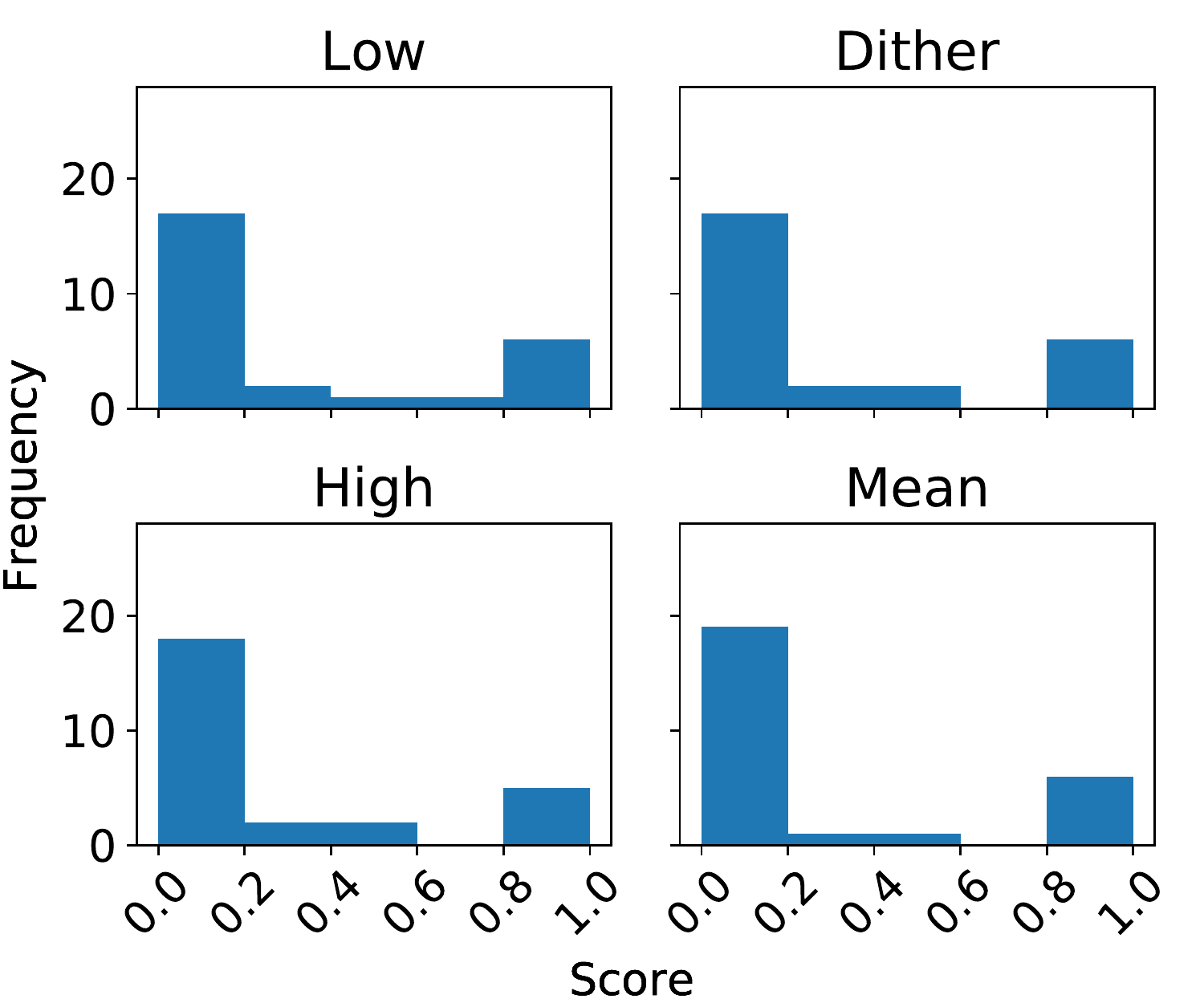}
        \caption{Test data.}
        \label{subfig:cnn_test_histo}
    \end{subfigure}%
    \vspace{-2mm}
    \caption{Distribution of the $\kappa$-statistic score ($\kappa_n$) on a per-shot basis, for the CNN.}    \label{fig:cnn_histograms}
\end{figure}

We trained the CNN for 250 epochs, allowing for the loss function to stabilize; each epoch consisted in 32 batches, with each batch containing 64 data samples. Upon completion of training, we tested the CNN's accuracy on both the training and test data. The model's results on ELM classification (ROC curve) can be seen in Figure \ref{fig:cnn_roc}. Table \ref{tab:kappa_cnn} shows the scores $\kappa_n$ and $\kappa_l$ for the entire dataset, while Figure \ref{fig:cnn_histograms} contains histograms showing the $\kappa_n$s distribution on a per-shot basis. 

\begin{table}[h!]
    \begin{center}
        \begin{tabular}{cccccc}
         &  & L & D & H & Mean \\
         \hline
        \multirow{2}{*}{$K_n$} & Train & 0.691 & 0.358 & 0.657 & 0.649 \\
         & Test & 0.219 & 0.115 & 0.157 & 0.182 \\
         \hline
        \multirow{2}{*}{$K_l$} & Train & 0.937 & 0.896 & 0.987 & 0.958 \\
         & Test & 0.941 & 0.848 & 0.986 & 0.962
        \end{tabular}
    \vspace{-3mm}
    \caption{$\kappa$-statistic scores ($\kappa_n$ and $\kappa_l)$ for each plasma mode and as a mean, on training and test data (values across all shots), for the CNN}
    \label{tab:kappa_cnn}
    \end{center}
\end{table}

\vspace{-5mm}

\newpage

\subsection{Conv-LSTM}

\begin{figure}[h!]
    \centering
    \begin{subfigure}{0.3\textwidth}
        \centering
        
        \includegraphics[width=\textwidth,
        scale = 0.6, 
        trim= 0 0 0 0,
        clip]{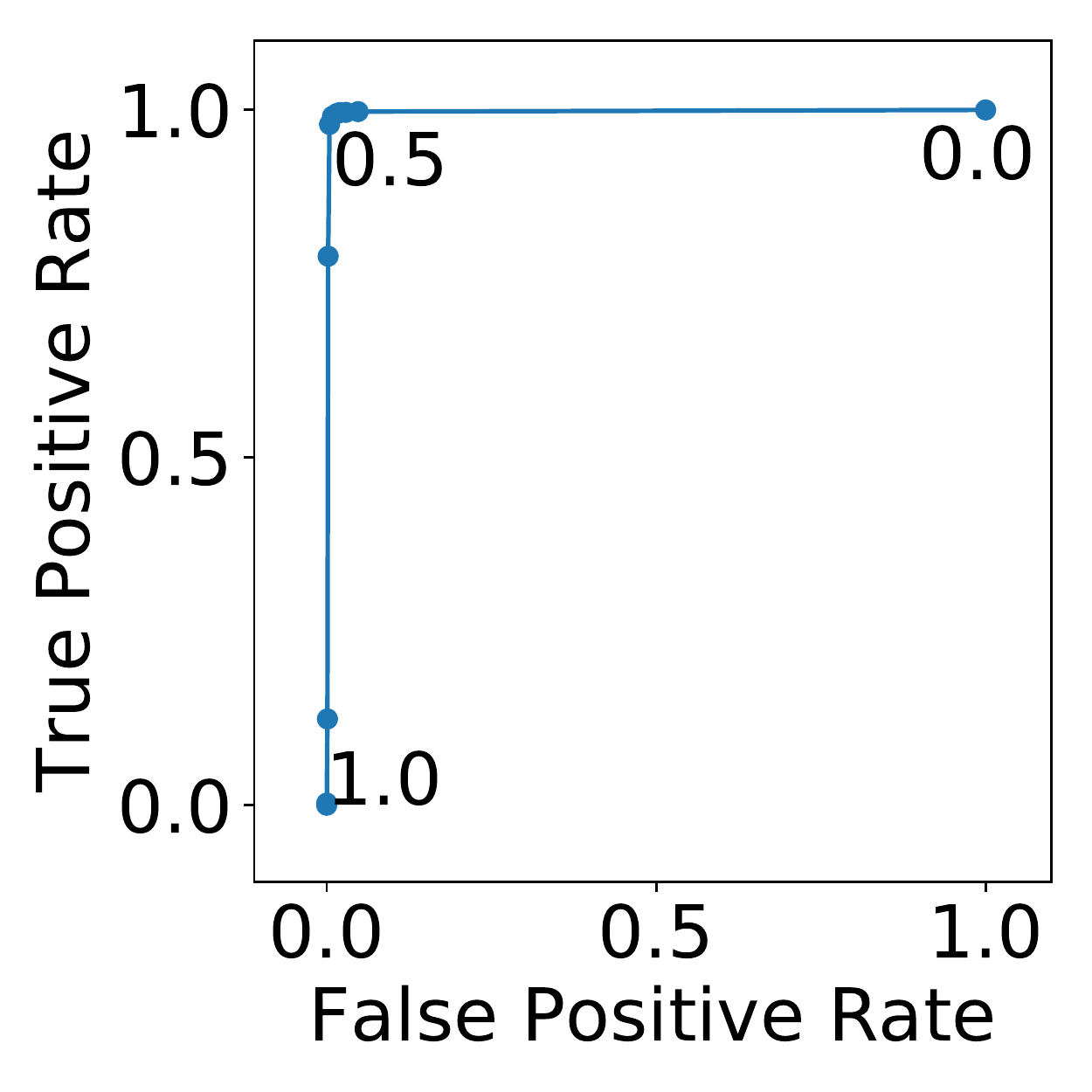}
        \caption{Training data.}
        \label{subfig:lstm_ROC_train}
    \end{subfigure}%
    ~
    \begin{subfigure}{0.3\textwidth}
        \centering
        
        \includegraphics[width=\textwidth,
        scale = 0.6, 
        trim= 0 0 0 0,
        clip]{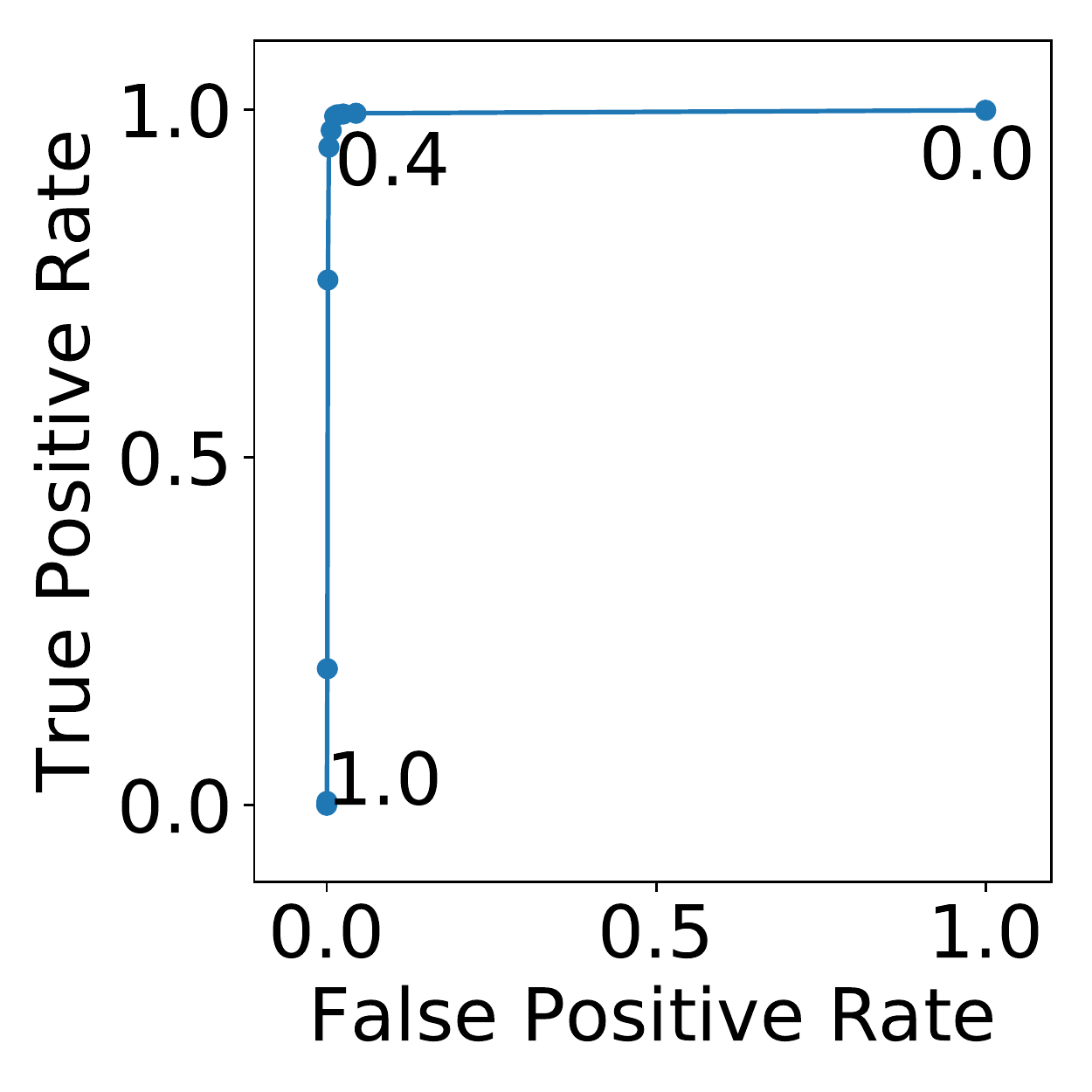}
        \caption{Test data.}
        \label{subfig:lstm_ROC_test}
    \end{subfigure}%
    \vspace{-3mm}
    \caption{ROC curves for ELM detection for the Conv-LSTM model. The detection threshold which maximizes the Youden index is $0.5$ for training and $0.4$ for test data; this yields index values of $0.977$ and $0.969$ for each set respectively. Using the ideal threshold for the training data ($0.5$) on the test data gives a slightly lower Youden index of $0.95$.}
    \label{fig:lstm_roc}
\end{figure}

\begin{figure}[h!]
    \centering
    \begin{subfigure}{0.45\textwidth}
        \centering
        
        \includegraphics[width=\textwidth,
        scale = 0.55, 
        trim= 0 0 0 0,
        clip]{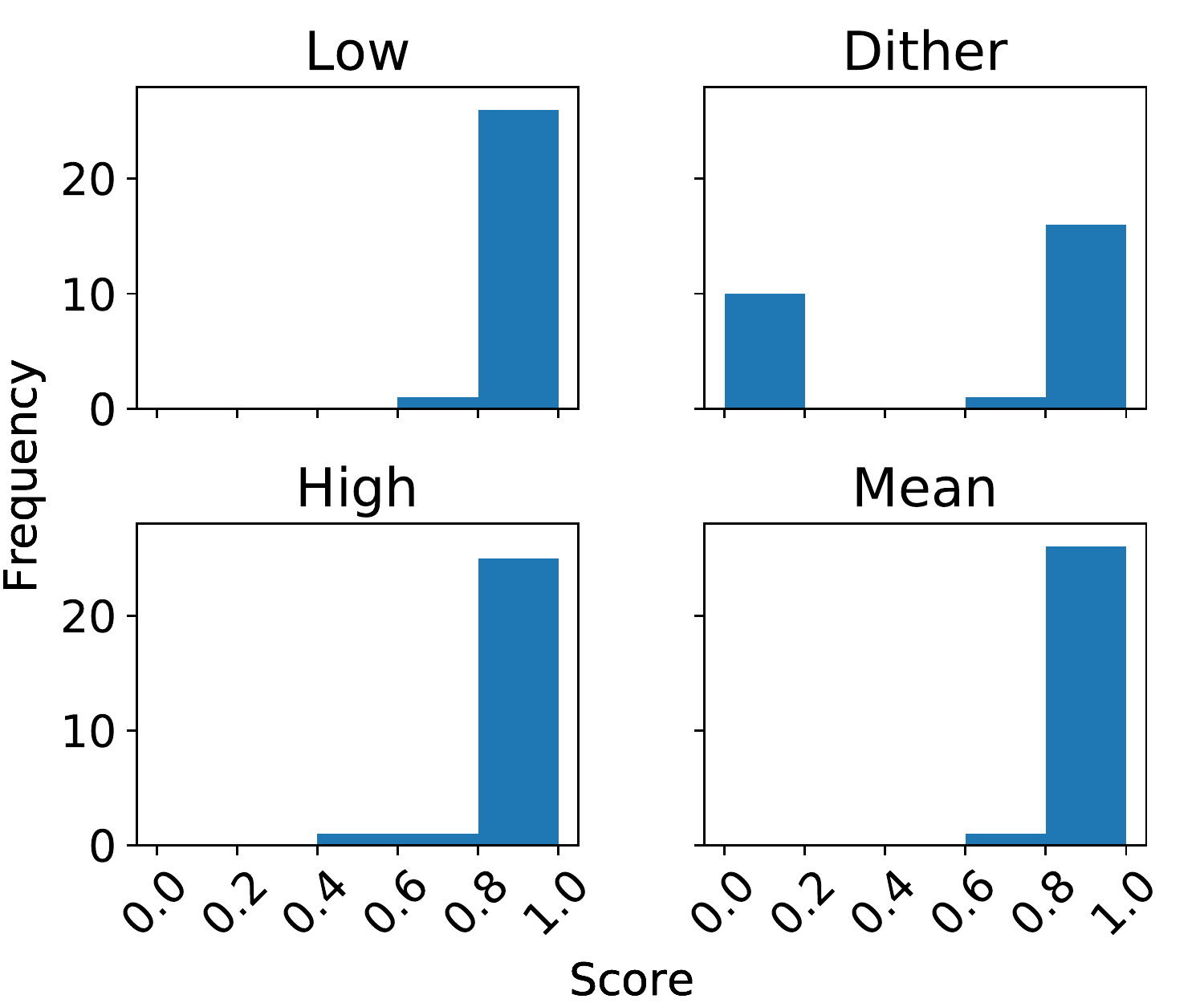}
        \caption{Training data.}
        \label{subfig:lstm_train_histo}
    \end{subfigure}%
    ~
    \begin{subfigure}{0.45\textwidth}
        \centering
        
        \includegraphics[width=\textwidth,
        scale = 0.55, 
        trim= 0 0 0 0,
        clip]{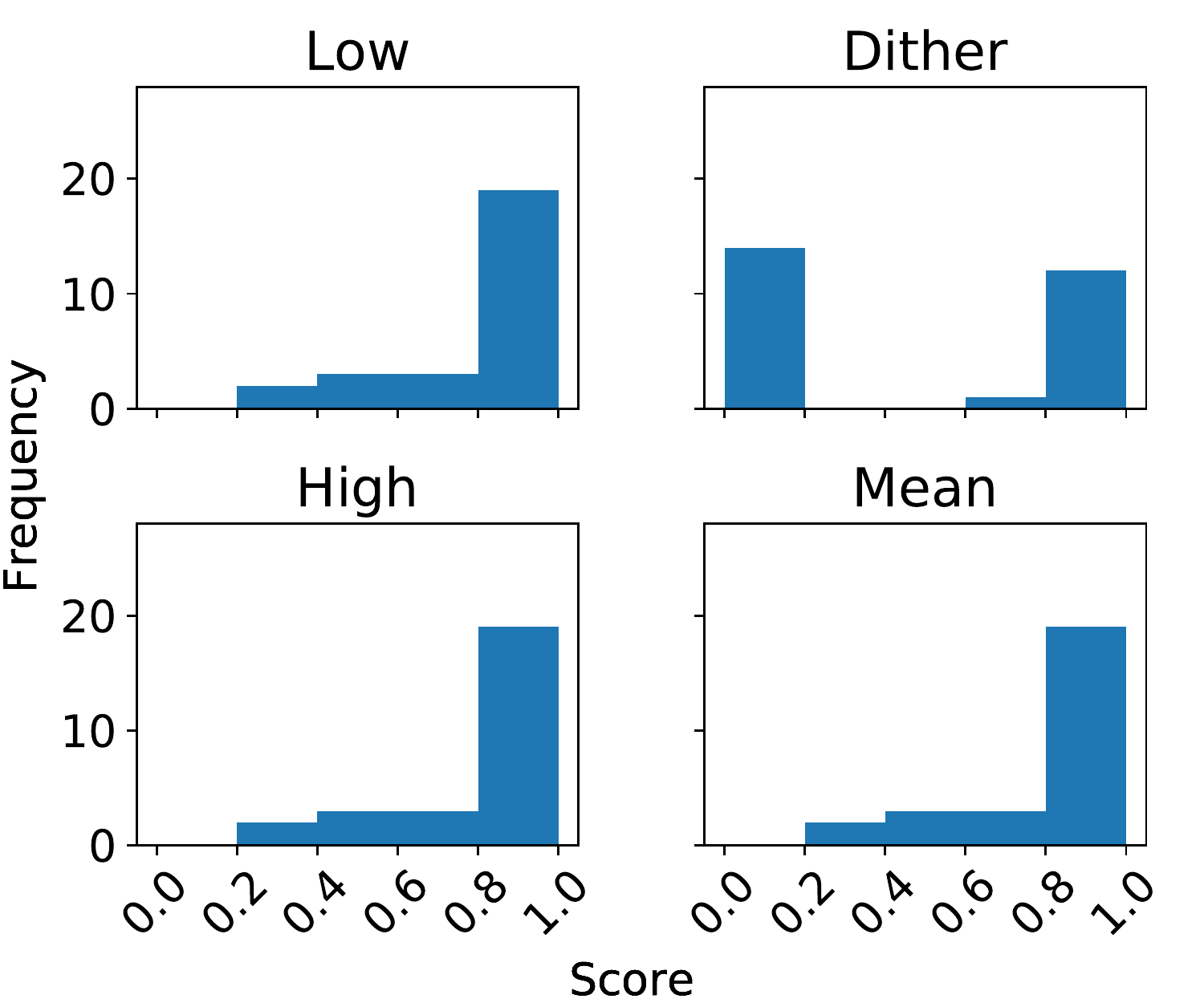}
        \caption{Test data.}
        \label{subfig:lstm_test_histo}
    \end{subfigure}%
    \caption{Distribution of the $\kappa$-statistic score ($\kappa_n$) on a per-shot basis, for the Conv-LSTM}
    \vspace{-7mm}
    \label{fig:lstm_histograms}
\end{figure}

We trained the convolutional LSTM for 400 epochs, allowing the loss function to stabilize. Each epoch consisted of 64 batches, with each batch containing 64 data samples. The results of computing scores $\kappa_l$ and $\kappa_n$, using the same definitions as for the CNN can be seen in tables \ref{tab:kappa_lstm}. The ROC curves detailing the results on ELM detection can be seen in Figure \ref{fig:lstm_roc}.  Figure \ref{fig:lstm_histograms} contains histograms showing the score $K_n$ values on a per-shot basis. 

\begin{table}[h!]
    \begin{center}
        \begin{tabular}{cccccc}
         &  & L & D & H & Mean \\
         \hline
        \multirow{2}{*}{$K_n$} & Train & 0.96 & 0.889 & 0.967 & 0.96 \\
         & Test & 0.82 & 0.766 & 0.85 & 0.832 \\
         \hline
        \multirow{2}{*}{$K_l$} & Train & 0.96 & 0.94 & 0.992 & 0.98 \\
         & Test & 0.901 & 0.808 & 0.98 & 0.935
        \end{tabular}
    \end{center}
    \vspace{-5mm}
    \caption{$\kappa$-statistic scores ($\kappa_n$ and $\kappa_l$) for each plasma mode on training and test data, for the Conv-LSTM  }
    \label{tab:kappa_lstm}
\end{table}
\vspace{-5mm}

\clearpage
\subsection{Discussion}

A comparison of the $\kappa_n$ scores on training and test data for each classifier shows that the vonvolutional LSTM performs better than the CNN for all three plasma states. Furthermore, looking at the distribution of the mean $\kappa_n$ scores on a per-shot basis through the histograms, one can see that the worst Conv-LSTM classifications do not have a score lower than 0.6 on training data, while for the CNN alone, even on training data, mean $\kappa_n$ scores lower than 0.2 exist. For both classifiers, the performance on training data surpasses that on test data, both on a state-by-state basis, and as a mean across all states, which indicates the occurrence of overfitting.

For both networks, an analysis of the $\kappa_l$ scores of their training and test data indicates that human labeler disagreement is highest for dithers -- the scores for that particular state are consistently lower. Interestingly, both networks also score their lowest results for dithers.

Comparing the Conv-LSTM's $\kappa_l$ and $\kappa_n$ scores shows that, at least on training data, the network behaves, on average, similarly to a single human labeler, making errors (or disagreeing with the ground truth) at approximately the same rate -- the mean $\kappa_l$ score for training data is 0.98, while the mean $\kappa_n$ score for training data is 0.96. On test data, the Conv-LSTM performs slightly worse than a single human labeler, as seen by the fact that the network's mean K-index score on test data $\kappa_n$ is 0.832, while $\kappa_l$ is 0.935. 

As measured by the Youden index, we achieve excellent performance in detection of ELMs on both training and test data using both models; the ideal detection thresholds generate true positive detection rates very close to 1, while bringing false positive detection rates essentially to 0.  The Youden indexes for test data are only slightly lower than for training data, which suggests that overfitting is minimal. Furthermore, for both models, on both training and test data, the ROC curves' points are mostly concentrated close to True Positive Rates of 1 and False Positive Rates of 0, which indicates that the choice of ELM detection threshold does not significantly change the behavior of the classifiers.

Finally, the scores for ELMs being essentially the same for both models indicates that the features in the data which allow for identification of ELMs are mostly local: the CNN, even without knowledge of long-term temporal correlations, performs excellent classification. 

Because the Conv-LSTM has highest $\kappa_n$ scores, we made a case-by-case analysis of that network's classification of all our available shots. Broadly, the Conv-LSTM's results on state detection, on a per-shot basis, can be placed into six different categories:

\begin{enumerate}
    \item A (sometimes very) short detection, of a dither that is not labeled in the data. Due to the way the K-score $\kappa_n$ is computed, a mistaken dither classification by the network of a single time point (in a whole sequence), in a shot which has no regions where the ground truth ($C_4$) is dithering, will bring the score for that state down to 0, even if the remainder of the shot is correctly classified (17 shots). 
    \item A clearly incorrect classification, of a temporal region of a shot as being in a dithering state (4 shots);
    \item A missed detection of an L-H transition (1 shot);
    \item A missed detection of an H-L transition (2 shots);
    \item An overall bad detection across an entire shot (7 shots);
    \item An overall good detection across an entire shot (23 shots).
\end{enumerate}

\begin{figure}[!htb]
    \centering
    \includegraphics[scale=0.39, trim=0 2 0 5, clip]{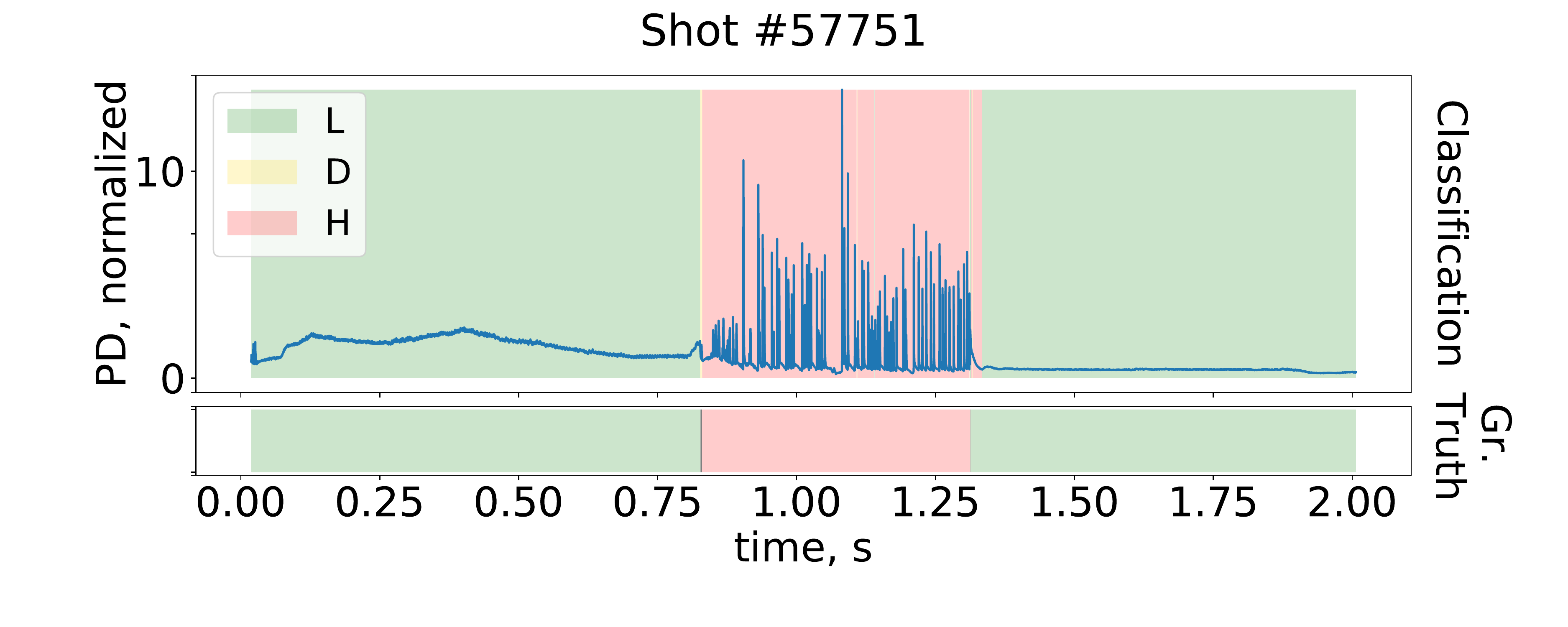}
    \vspace{-6mm}
    \caption{TCV shot \#57751 (PD signal) and the Conv-LSTM's classification of state as the shot evolves. Notice the (very short) detected dithering phase shortly after $t=0.75$: no dithers are present in the labels, so the score for D is 0.}
    \vspace{-7mm}
    \label{fig:57751_results}
\end{figure}

\begin{figure}[!htb]
    \centering
    \includegraphics[scale=0.39, trim=0 2 0 5, clip]{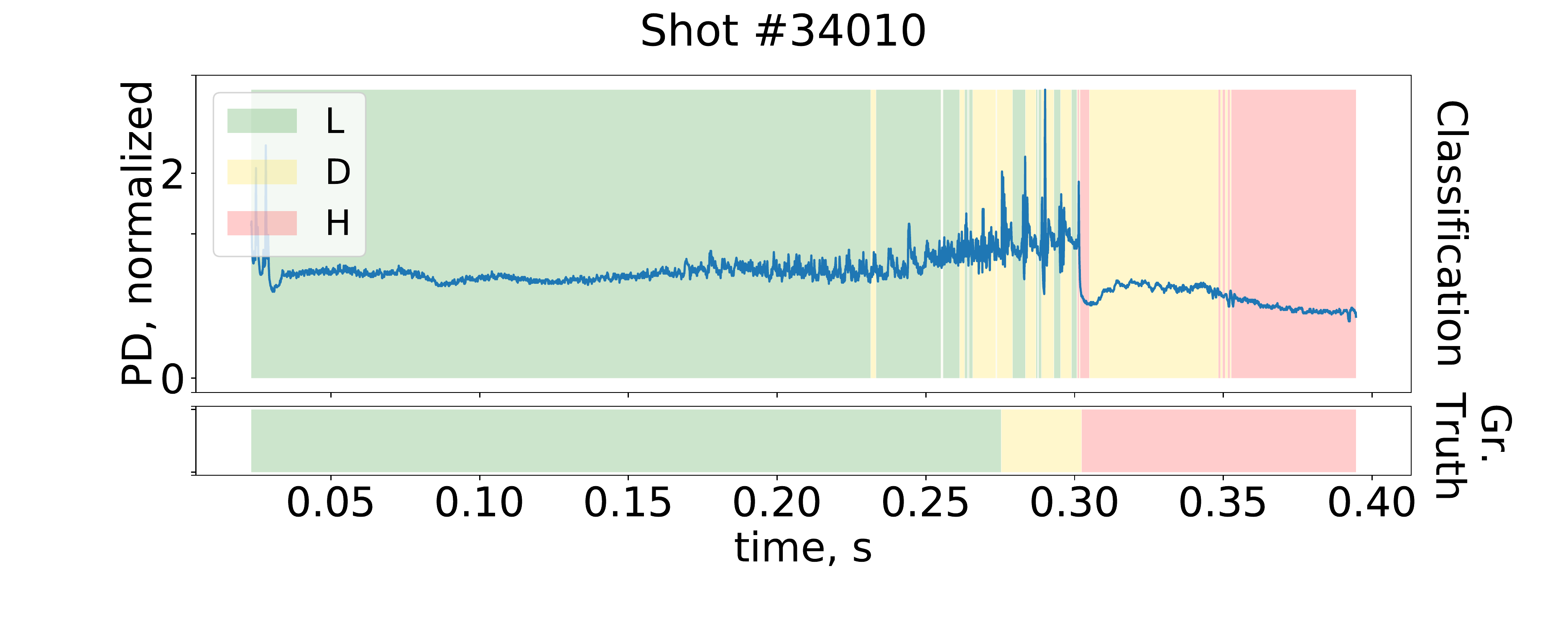}
    \vspace{-6mm}
    \caption{In TCV shot \# 34010, the network correctly identifies the transition into H mode at $t=0.3s$, but it shortly thereafter (incorrectly) switches back to dithering. }
    \label{fig:34010_results}
\end{figure}

\begin{figure}[!htb]
    \centering
    \includegraphics[scale=0.39, trim=0 2 0 5, clip]{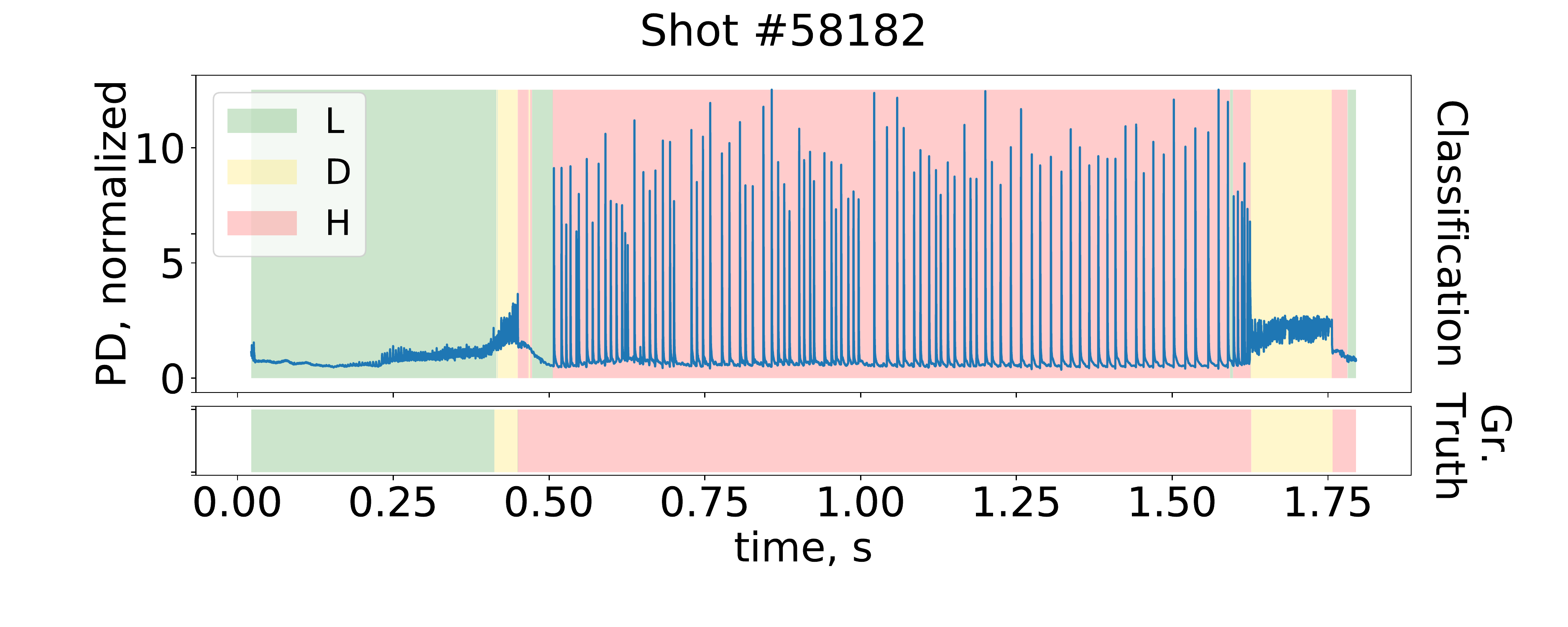}
    \vspace{-6mm}
    \caption{In TCV shot \#58182, the network correctly identifies a transition into H mode (shortly before $t=0.5s$) but then incorrectly switches back to L mode and remains there until the first ELMs (spikes in the PD signal) appear. }
    \label{fig:58182_results}
\end{figure}

\begin{figure}[!htb]
    \centering
    \includegraphics[scale=0.39, trim=0 2 0 5, clip]{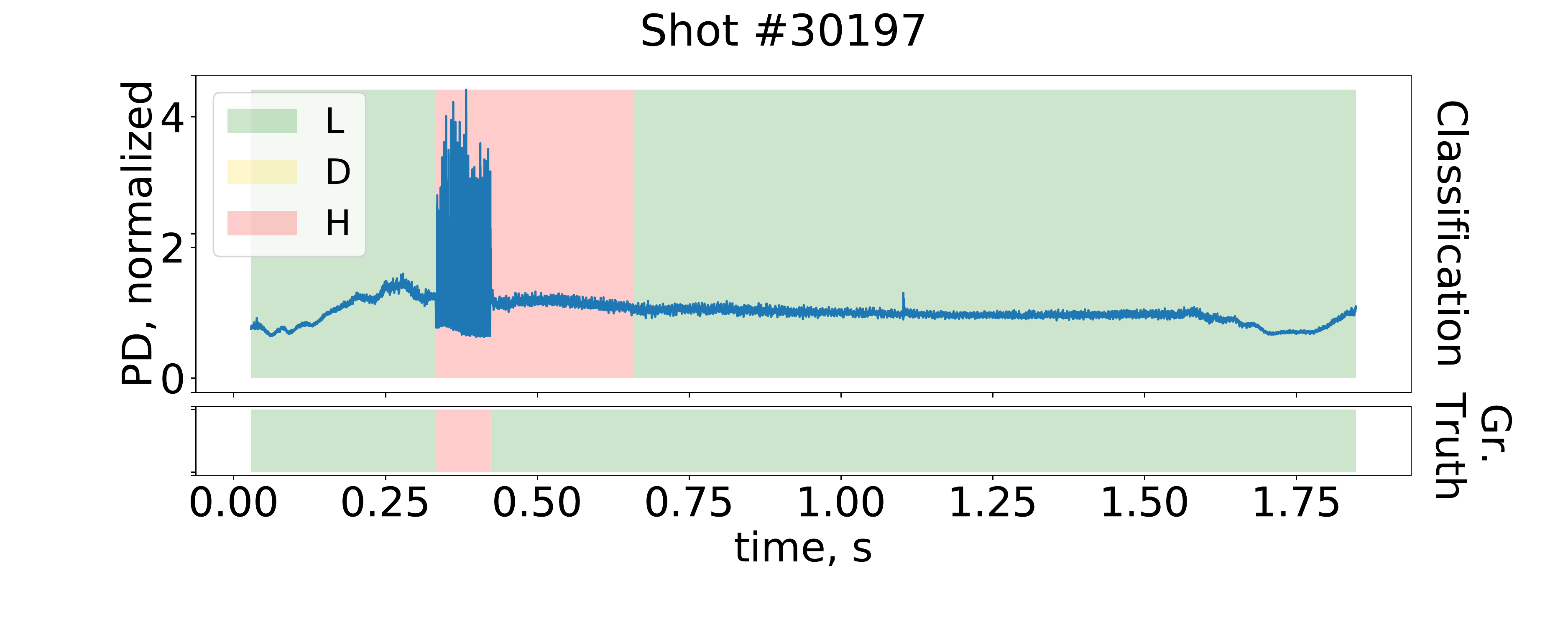}
    \vspace{-6mm}
    \caption{In shot \#30197, the network misses the transition from H to L mode, which happens immediately after the series of spikes in the PD signal, and only makes the switch after $t=0.5s$.}
    \label{fig:30197_results}
\end{figure}

\begin{figure}[!htb]
    \centering
    \includegraphics[scale=0.39, trim=0 2 0 5, clip]{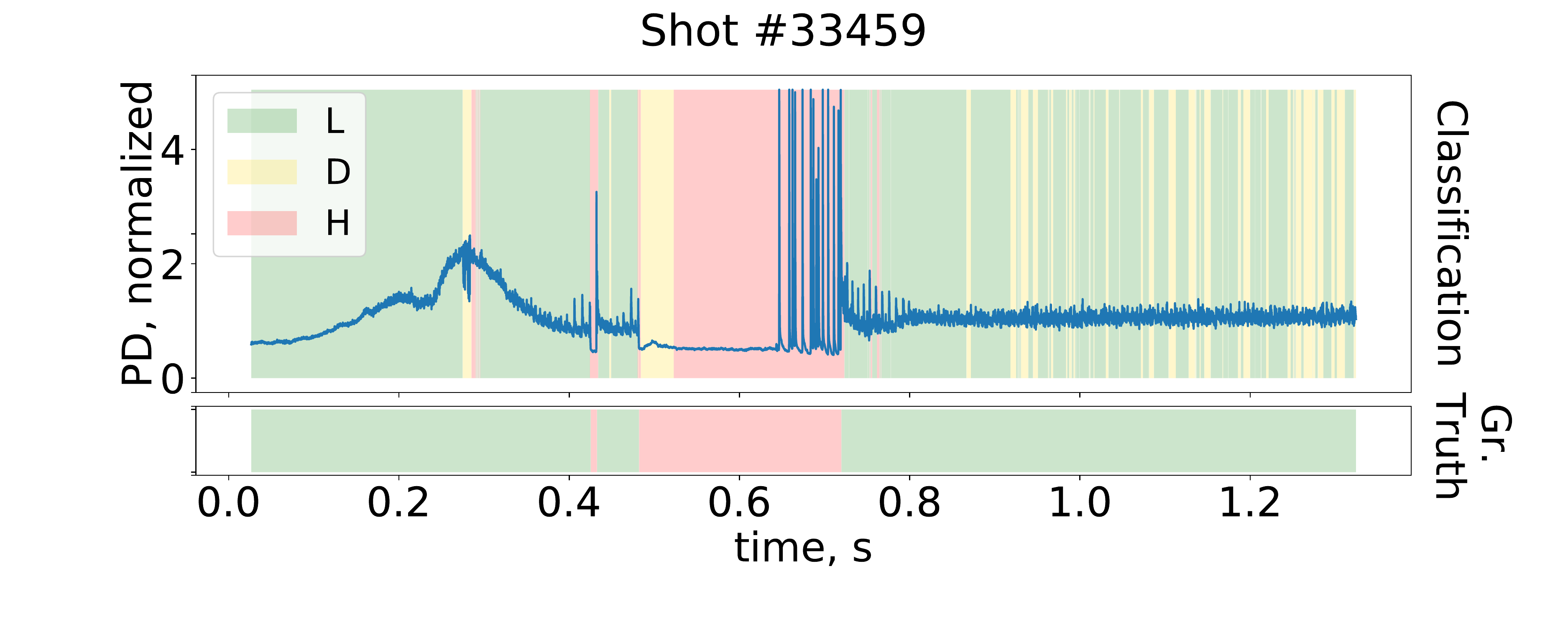}
    \vspace{-6mm}
    \caption{Shot \#33459 represents an overall bad classification by the network; many dithers are incorrectly classified, while the transition from L to H mode is missed. Around $t=0.3s$, immediately after classifying a D mode, the network oscillates between L and H in quick succession for about $0.01s$, which to the naked eye might appear in this plot as a gray area; in reality, it is an artifact of the plot, with alternating red and green regions.  }
    \label{fig:33459_results}
\end{figure}

\begin{figure}[!htb]
    \centering
    \includegraphics[scale=0.39, trim=0 2 0 5, clip]{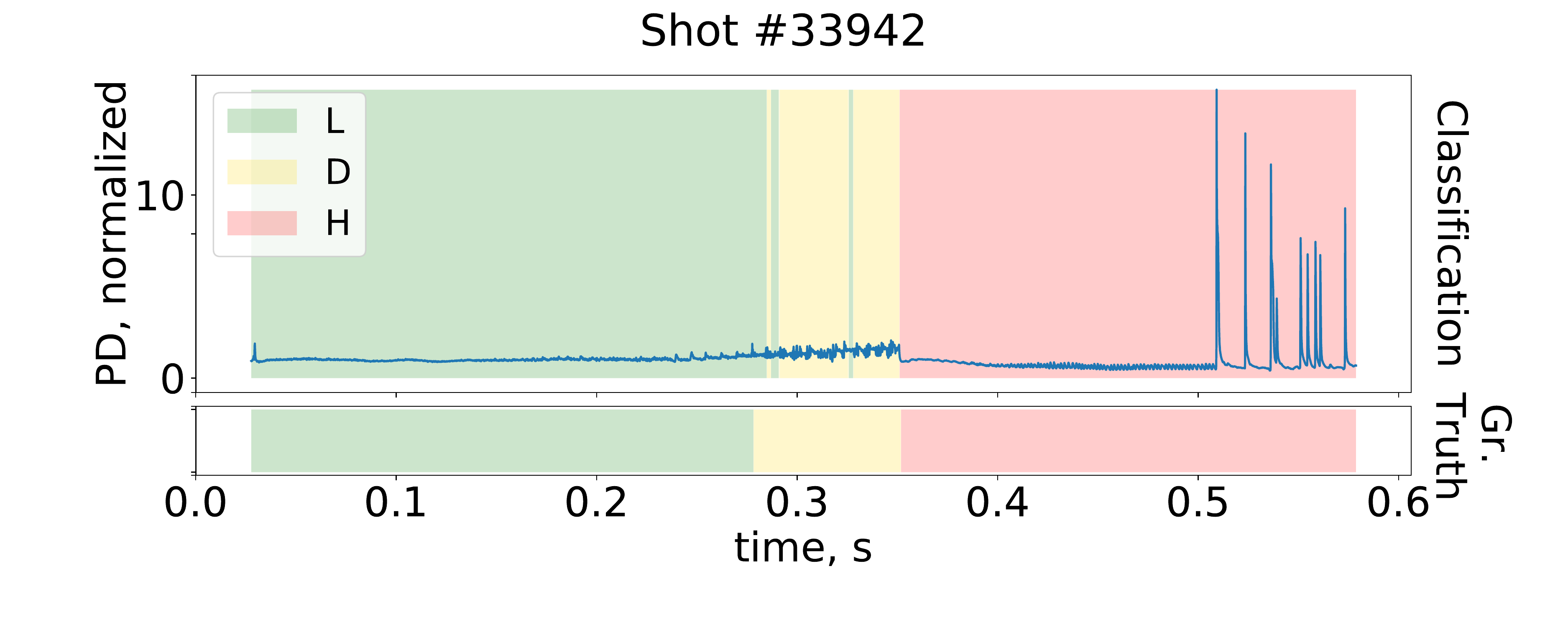}
    \vspace{-6mm}
    \caption{Shot \#33942 is an example of an overall good detection. }
    \label{fig:33942_results}
\end{figure}

Table \ref{tab:clstm_per_shot} lists 6 shots which are representative of each of the types of results listed above. The table shows the computed $\kappa_n$ scores for each of those shots on a per-state basis, as well as the score's mean value, and the fraction of time, for the ground truth of each shot, that a particular state is labeled. The table also lists which of the 6 cases above the shot is representative of. Figures \ref{fig:57751_results} to \ref{fig:33942_results} are plots of those same shots, where the background color in the top plot denotes the state detected by the Conv-LSTM, and in the bottom plot, denotes the ground truth label. Small gray areas in the bottom plot denote regions where ground truth is not defined, i.e., there is no majority agreement between labelers. 

\begin{table}[h]

    \begin{center}
        \item[]
        \begin{tabular}{|c|c|c|c|c|c|c|c|c|}
        \hline
        \multirow{2}{*}{Case} & \multirow{2}{*}{Shot ID} & \multicolumn{2}{c|}{L} & \multicolumn{2}{c|}{D} & \multicolumn{2}{c|}{H} & \multirow{2}{*}{Mean} \\ \cline{3-8}
         &  & Fraction & Score & Fraction & Score & Fraction & Score &  \\ \hline
        1 & 57751 & 0.756 & 0.97 & 0 & 0 & 0.243 & 0.97 & 0.97 \\ \hline
        2 & 34010 & 0.679 & 0.856 & 0.073 & 0.232 & 0.248 & 0.602 & 0.748 \\ \hline
        3 & 58182 & 0.22 & 0.912 & 0.095 & 0.969 & 0.685 & 0.927 & 0.928 \\ \hline
        4 & 30197 & 0.951 & 0.384 & 0 & 1 & 0.049 & 0.384 & 0.384 \\ \hline
        5 & 33459 & 0.811 & 0.662 & 0 & 0 & 0.189 & 0.846 & 0.697 \\ \hline
        6 & 33942 & 0.455 & 0.953 & 0.183 & 0.884 & 0.412 & 0.997 & 0.962 \\ \hline
        \end{tabular}

    \end{center}
    \vspace{-5mm}
    \caption{Kappa statistic ($\kappa_n$) scores for each plasma mode on training and test data for selected shots representative of each of the six result categories }
    \label{tab:clstm_per_shot}
\end{table}

\clearpage

\section{Conclusions} \label{sec:conclusions}

We have developed two Deep Learning-based classifiers to perform automatic detection of ELMs and classification of plasma modes. The task was two-fold: on one hand, to perform a binary classification, for each time slice of a plasma shot, on whether an ELM is occurring or not; and, to automatically determine which plasma mode (or alternatively, whether a transition between plasma modes) is occurring. One approach is to use a convolutional Neural Network (CNN), which uses only local correlations in data to perform classification. The second approach uses a Convolutional LSTM (Conv-LSTM) neural network, which also takes advantages of long-term temporal correlations in data. 

On ELM detection, the two networks can achieve essentially equal results. On the plasma state classification, a clear difference can be seen between the results obtained with the CNN, and those obtained with the Conv-LSTM. Comparing the $\kappa$-index ($\kappa_n$) scores of each network shows that the LSTM's scores are clearly higher, which suggests that, at least when it comes to detection of plasma modes, the processing of long-term correlations in data facilitates accurate classification. 
There is some indication that overfitting occurred. However, our monitoring of the training progression indicated that, while the metric values for test data are always lower, they did, nevertheless, become better as training progressed. Thus, an overfitting-avoidance strategy such as early stopping would, in this case, not have helped achieve better test accuracy.

While the results from the Conv-LSTM are better, that network is also more complex with both network training and inference taking longer. 

Although this work used data from the TCV tokamak, it should also be possible to adapt it to other machines; as a matter of fact, the data sources used exist on most tokamaks. As long as the data fed to the neural networks is from those same sources, this model could in principle be used for automatic labeling of shots from a number of different machines.

\section*{Acknowledgements}
\small
\noindent This work has been carried out within the framework of the EUROfusion Consortium and has received funding from the Euratom research and training programme 2014-2018 and 2019-2020 under grant agreement No 633053. The views and opinions expressed herein do not necessarily reflect those of the European Commission. We would like express our gratitude to B. Labit, R. Maurizio and O. Sauter at SPC/EPFL for taking the time to manually label the data used for training. This work was supported in part by the Swiss National Science Foundation.

\clearpage

\section*{References}
\bibliography{iopart-num}

\clearpage

\end{document}